\documentclass[%
reprint, 
superscriptaddress,
amsmath,amssymb,
]{revtex4-2}

\usepackage{graphicx}
\usepackage{dcolumn}
\usepackage{bm}
\usepackage{hyperref}
\usepackage[mathlines]{lineno}

\usepackage{xcolor}

\begin{document}

\preprint{???}

\title{\textbf{
Identification of rippling interface between laser-ablated plasma and ambient uniform plasma at rest
}
}%

\author{R.~Yamazaki}
\affiliation{Department of Physical Sciences, Aoyama Gakuin University, 5-10-1 Fuchinobe, Sagamihara, Kanagawa 252-5258, Japan}
\email[]{ryo@phys.aoyama.ac.jp (R.Y.)}
\affiliation{Institute of Laser Engineering, Osaka University, 2-6, Yamadaoka, Suita, Osaka 565-0871, Japan}

\author{S.~Tomita}
\affiliation{Institute for Advanced Academic Research, Chiba University, 1-33 Yayoi-cho, Inage-ku, Chiba 263-8522, Japan}
\affiliation{Laboratory of Physics, College of Science and Technology, Nihon University, Funabashi, Chiba 274-8501, Japan}

\author{T.~Takezaki}
\affiliation{Faculty of Engineering, University of Toyama, 3190, Gofuku, Toyama 930-8555, Japan}

\author{T.~Morita}
\affiliation{Faculty of Engineering Sciences, Kyushu University, 6-1 Kasuga-Koen, Kasuga, Fukuoka 816-8580, Japan}

\author{S. J. Tanaka}
\affiliation{Department of Physical Sciences, Aoyama Gakuin University, 5-10-1 Fuchinobe, Sagamihara, Kanagawa 252-5258, Japan}
\affiliation{Frontier Research Institute for Interdisciplinary Sciences, Tohoku University, 6-3 Aramaki Aza-Aoba, Aoba-ku, Sendai 980-8578, Japan}

\author{Y.~Ohira}
\affiliation{Department of Earth and Planetary Science, The University of Tokyo, 7-3-1 Hongo, Bunkyo, Tokyo 113-0033, Japan}

\author{T.~Umeda}
\affiliation{Information Initiative Center, Hokkaido University, Sapporo 060-0811, Japan}


\author{K.~Aihara}
\affiliation{Department of Physical Sciences, Aoyama Gakuin University, 5-10-1 Fuchinobe, Sagamihara, Kanagawa 252-5258, Japan}

\author{S.~Kambayashi}
\affiliation{Department of Physical Sciences, Aoyama Gakuin University, 5-10-1 Fuchinobe, Sagamihara, Kanagawa 252-5258, Japan}

\author{R.~Kawamura}
\affiliation{Department of Physical Sciences, Aoyama Gakuin University, 5-10-1 Fuchinobe, Sagamihara, Kanagawa 252-5258, Japan}

\author{K.~Obayashi}
\affiliation{Department of Physical Sciences, Aoyama Gakuin University, 5-10-1 Fuchinobe, Sagamihara, Kanagawa 252-5258, Japan}

\author{M.~Murakoshi}
\affiliation{Department of Physical Sciences, Aoyama Gakuin University, 5-10-1 Fuchinobe, Sagamihara, Kanagawa 252-5258, Japan}

\author{Y.~Sato}
\affiliation{Department of Physical Sciences, Aoyama Gakuin University, 5-10-1 Fuchinobe, Sagamihara, Kanagawa 252-5258, Japan}

\author{R.~Shiina}
\affiliation{Department of Physical Sciences, Aoyama Gakuin University, 5-10-1 Fuchinobe, Sagamihara, Kanagawa 252-5258, Japan}

\author{A.~Takata}
\affiliation{Department of Physical Sciences, Aoyama Gakuin University, 5-10-1 Fuchinobe, Sagamihara, Kanagawa 252-5258, Japan}

\author{K.~Matsui}
\affiliation{Department of Physical Sciences, Aoyama Gakuin University, 5-10-1 Fuchinobe, Sagamihara, Kanagawa 252-5258, Japan}

\author{S.~Ueno}
\affiliation{Department of Physical Sciences, Aoyama Gakuin University, 5-10-1 Fuchinobe, Sagamihara, Kanagawa 252-5258, Japan}

\author{Y.~Kidokoro}
\affiliation{Department of Physical Sciences, Aoyama Gakuin University, 5-10-1 Fuchinobe, Sagamihara, Kanagawa 252-5258, Japan}

\author{S.~Yakura}
\affiliation{Department of Physical Sciences, Aoyama Gakuin University, 5-10-1 Fuchinobe, Sagamihara, Kanagawa 252-5258, Japan}

\author{Y.~Okada}
\affiliation{Department of Physical Sciences, Aoyama Gakuin University, 5-10-1 Fuchinobe, Sagamihara, Kanagawa 252-5258, Japan}

\author{S.~Suzuki}
\affiliation{Department of Physical Sciences, Aoyama Gakuin University, 5-10-1 Fuchinobe, Sagamihara, Kanagawa 252-5258, Japan}

\author{H.~Kondo}
\affiliation{Department of Physical Sciences, Aoyama Gakuin University, 5-10-1 Fuchinobe, Sagamihara, Kanagawa 252-5258, Japan}

\author{N.~Shimoda}
\affiliation{Department of Physical Sciences, Aoyama Gakuin University, 5-10-1 Fuchinobe, Sagamihara, Kanagawa 252-5258, Japan}

\author{J.~Shiota}
\affiliation{Department of Physical Sciences, Aoyama Gakuin University, 5-10-1 Fuchinobe, Sagamihara, Kanagawa 252-5258, Japan}

\author{Y.~Sudo}
\affiliation{Department of Physical Sciences, Aoyama Gakuin University, 5-10-1 Fuchinobe, Sagamihara, Kanagawa 252-5258, Japan}

\author{D.~Suzuki}
\affiliation{Department of Physical Sciences, Aoyama Gakuin University, 5-10-1 Fuchinobe, Sagamihara, Kanagawa 252-5258, Japan}

\author{K.~Sugiyama}
\affiliation{Department of Physical Sciences, Aoyama Gakuin University, 5-10-1 Fuchinobe, Sagamihara, Kanagawa 252-5258, Japan}

\author{A.~Morita}
\affiliation{Department of Physical Sciences, Aoyama Gakuin University, 5-10-1 Fuchinobe, Sagamihara, Kanagawa 252-5258, Japan}

\author{H.~Kawakatsu}
\affiliation{Department of Physical Sciences, Aoyama Gakuin University, 5-10-1 Fuchinobe, Sagamihara, Kanagawa 252-5258, Japan}

\author{A.~Kashima}
\affiliation{Department of Physical Sciences, Aoyama Gakuin University, 5-10-1 Fuchinobe, Sagamihara, Kanagawa 252-5258, Japan}

\author{Y.~Ishimura}
\affiliation{Department of Physical Sciences, Aoyama Gakuin University, 5-10-1 Fuchinobe, Sagamihara, Kanagawa 252-5258, Japan}

\author{K.~Umehara}
\affiliation{Department of Physical Sciences, Aoyama Gakuin University, 5-10-1 Fuchinobe, Sagamihara, Kanagawa 252-5258, Japan}

\author{S.~Egashira}
\affiliation{Graduate School of Science, Osaka University, 1-1 Machikane-yama, Toyonaka, Osaka 560-0043, Japan}

\author{H.~Ishihara}
\affiliation{Graduate School of Science, Osaka University, 1-1 Machikane-yama, Toyonaka, Osaka 560-0043, Japan}

\author{O.~Kuramoto}
\affiliation{Graduate School of Science, Osaka University, 1-1 Machikane-yama, Toyonaka, Osaka 560-0043, Japan}

\author{Y.~Matsumoto}
\affiliation{Graduate School of Science, Osaka University, 1-1 Machikane-yama, Toyonaka, Osaka 560-0043, Japan}

\author{K.~Maeda}
\affiliation{Graduate School of Science, Osaka University, 1-1 Machikane-yama, Toyonaka, Osaka 560-0043, Japan}

\author{Y.~Suzuki}
\affiliation{Graduate School of Science, Osaka University, 1-1 Machikane-yama, Toyonaka, Osaka 560-0043, Japan}

\author{M.~Hanano}
\affiliation{Graduate School of Science, Osaka University, 1-1 Machikane-yama, Toyonaka, Osaka 560-0043, Japan}

\author{Y.~Toyoda}
\affiliation{Graduate School of Science, Osaka University, 1-1 Machikane-yama, Toyonaka, Osaka 560-0043, Japan}

\author{K.~Kuramoto}
\affiliation{Graduate School of Engineering, Osaka University, 2-1, Yamadaoka, Suita, Osaka 565-0871, Japan}

\author{F.~Nikaido}
\affiliation{Graduate School of Engineering, Osaka University, 2-1, Yamadaoka, Suita, Osaka 565-0871, Japan}

\author{K.~Iida}
\affiliation{Graduate School of Engineering, Osaka University, 2-1, Yamadaoka, Suita, Osaka 565-0871, Japan}

\author{Y. Naito}
\affiliation{Graduate School of Engineering, Osaka University, 2-1, Yamadaoka, Suita, Osaka 565-0871, Japan}

\author{Y.~Sato}
\affiliation{Interdisciplinary Graduate School of Engineering Sciences, Kyushu University, Kasuga 816-8580, Japan}

\author{K.~Takahashi}
\affiliation{Interdisciplinary Graduate School of Engineering Sciences, Kyushu University, Kasuga 816-8580, Japan}

\author{K.~Nakahara}
\affiliation{Interdisciplinary Graduate School of Engineering Sciences, Kyushu University, Kasuga 816-8580, Japan}

\author{Y.~Maenosono}
\affiliation{Interdisciplinary Graduate School of Engineering Sciences, Kyushu University, Kasuga 816-8580, Japan}

\author{Y.~Muramoto}
\affiliation{Interdisciplinary Graduate School of Engineering Sciences, Kyushu University, Kasuga 816-8580, Japan}

\author{Y.~Kanesada}
\affiliation{Interdisciplinary Graduate School of Engineering Sciences, Kyushu University, Kasuga 816-8580, Japan}

\author{K.~Koba}
\affiliation{Interdisciplinary Graduate School of Engineering Sciences, Kyushu University, Kasuga 816-8580, Japan}

\author{T.~Ogawa}
\affiliation{Interdisciplinary Graduate School of Engineering Sciences, Kyushu University, Kasuga 816-8580, Japan}

\author{R.~Shiiba}
\affiliation{Interdisciplinary Graduate School of Engineering Sciences, Kyushu University, Kasuga 816-8580, Japan}

\author{T.~Oguchi}
\affiliation{Faculty of Engineering, University of Toyama, 3190, Gofuku, Toyama 930-8555, Japan}


\author{M.~Edamoto}
\affiliation{Faculty of Science and Technology, Seikei University, 3-3-1 Kichijoji-Kitamachi, Musashino, Tokyo, 180-8633, Japan}

\author{M.~Iwamoto}
\affiliation{Graduate School of System Informatics, Kobe University, 1-1 Rokkodai-cho, Nada, Kobe, Hyogo, 657-8501  Japan}

\author{M.~Hoshino}
\affiliation{Department of Earth and Planetary Science, The University of Tokyo, 7-3-1 Hongo, Bunkyo, Tokyo 113-0033, Japan}

\author{T.~Inoue}
\affiliation{Department of Physics, Konan University, Okamoto 8-9-1, Higashinada-ku, Kobe 658-8501, Japan}

\author{S.~Isayama}
\affiliation{Faculty of Engineering Sciences, Kyushu University, 6-1 Kasuga-Koen, Kasuga, Fukuoka 816-8580, Japan}

\author{A.~T.~Ishii}
\affiliation{Faculty of Science, Yamagata University, 1-4-12 Kojirakawa-Machi, Yamagata, Yamagata 990-8560, Japan}

\author{Y.~Kuramitsu}
\affiliation{Graduate School of Engineering, Osaka University, 2-1, Yamadaoka, Suita, Osaka 565-0871, Japan}

\author{S. Matsukiyo}
\affiliation{Faculty of Engineering Sciences, Kyushu University, 6-1 Kasuga-Koen, Kasuga, Fukuoka 816-8580, Japan}

\author{N.~Ohnishi}
\affiliation{Department of Aerospace Engineering, Tohoku University,  6-6 Aramaki Aza Aoba, Aoba, Sendai, Miyagi 980-8579, Japan}

\author{K.~Sakai}
\affiliation{National Institute for Fusion Science, 322-6 Oroshi-cho, Toki, Gifu 509-5292, Japan}

\author{T.~Sano}
\affiliation{Institute of Laser Engineering, Osaka University, 2-6, Yamadaoka, Suita, Osaka 565-0871, Japan}

\author{K.~Tomita}
\affiliation{Division of Quantum Science and Engineering, Hokkaido University, Sapporo 060-8628, Japan}

\author{Y.~Sakawa}
\affiliation{Institute of Laser Engineering, Osaka University, 2-6, Yamadaoka, Suita, Osaka 565-0871, Japan}


\date{\today}

\begin{abstract}

We present laboratory experiments on the early stage of the formation process of 
quasi-perpendicular supercritical magnetized collisionless shocks using laser-produced plasmas in an ambient uniform
nitrogen (N) gas with 1~Torr under the external magnetic field with controlled polarity.
An expanding aluminum (Al) piston plasma arising from laser irradiation interacts with a magnetized ambient
N plasma, producing a bright, asymmetric self-emission  whose
global morphology reverses with the magnetic-field polarity.
Two-dimensional ideal-magnetohydrodynamics simulations reproduce the main features of the observed
structures and indicate that the ripple formation is unlikely to be driven by
the Kelvin--Helmholtz instability, but is instead associated with vortex
formation at the rounded leading edge of the piston plasma.
We show that the bright, rippling emission delineates the Al-N interface, whose intrinsic width is consistent with an electron kinetic scale, 
and that its upstream side is predominantly composed of N plasma.
This identification allows the formation of a supercritical collisionless shock in the magnetized N plasma to be discussed with minimal Al contamination.

\end{abstract}

\maketitle


\section{INTRODUCTION}\label{sec:intro}

Collisionless shocks are ubiquitous in astrophysical objects such as supernova remnants, as well as in solar-terrestrial and laboratory plasmas
\cite{Balogh2013,Burgess2015}.
The kinetic energy of the incoming plasma is dissipated into 
various components such as high-temperature ions and electrons, magnetic turbulence, and nonthermal particles \cite{Marcowith2016}.
In particular, supercritical magnetized collisionless shocks are of great interest 
in connection with astrophysical shocks associated with particle acceleration.
The detailed mechanisms around 
magnetized collisionless shocks are still not well understood despite state-of-the-art observations
\cite{Bamba2003,Johlander2016,Madanian2021,Agapitov2023,Graham2025},
particle-in-cell (PIC) 
\cite{Scholer2003,Lembege2009,Matsukiyo2011,Matsumoto2013,Sironi2013,Umeda2014,Yamazaki2019,Bohdan2020,Law2025}
and hybrid
\cite{Kropotina2023,Orusa2023,Orusa2025,Orusa2026}
simulations.
Experiments using high-power lasers provide another approach to studying collisionless shocks because they can combine controlled plasma conditions with spatially and temporally resolved diagnostics. 
Meanwhile, for unmagnetized upstream plasmas, laser experiments have demonstrated collisionless-shock formation and electron acceleration \cite{Fiuza2020}.
Motivated by these experimental advances, recent PIC simulations and theoretical studies 
have advanced our understanding of electron heating and energy partition between electrons and ions 
in unmagnetized or weakly magnetized collisionless shocks and counterstreaming plasmas \cite{Vanthieghem2022,Vanthieghem2024,Marret2026}.
In magnetized upstream plasmas as well, 
comparably well-characterized shock experiments are therefore needed to clarify how collisionless shocks form and mediate the dissipation of the incoming flow energy.

In quasi-perpendicular supercritical magnetized collisionless shocks, where the external upstream magnetic field is nearly perpendicular to the shock normal,
it is well known that incoming ions are reflected by the shock potential
\cite{Woods1971,Leroy1983,Sckopke1983,Gosling1985,Burgess1989,Wilkinson1990,Lembege1992}.
Experimental verification of the presence of such reflected ions would therefore
provide strong evidence for the formation of supercritical magnetized collisionless
shocks.
A number of high-power laser experiments have reported the generation of
supercritical magnetized collisionless shocks
\cite{Schaeffer2017,Schaeffer2019,Yao2021,Campbell2024,Johnson2025,Tang2025,Valenzuela-Villaseca2026},
and some of them claimed observations of reflected ions.
However, whether these observations truly capture ions reflected by the shock
potential requires careful examination.
One of the main reasons is that, in usual experimental setups, the upstream magnetized plasma
is spatially non-uniform, making it difficult to experimentally characterize the
properties of the unshocked upstream plasma with sufficient accuracy.
In particular, during the early stage of shock formation, the interface between the
ambient magnetized plasma and the laser-ablated piston plasma is often not yet fully
separated from the forming shock within the magnetized plasma.
Under such circumstances, it becomes challenging to distinguish whether the observed
reflected ions originate from reflection at the shock potential in the shock
transition layer or from reflection at the interface itself.
In the companion paper \cite{Yamazaki2025a}, we discuss reflected ions in a foot-like region and the associated ion temperature anisotropy in a forming shock. For such an interpretation, however, it is first essential to determine where the interface between the piston and ambient plasmas is located relative to the forming shock structure.

In several previous high-power-laser experiments on quasi-perpendicular supercritical magnetized collisionless-shock formation, pronounced density structures associated with the piston--ambient interaction, shock precursors, and shock formation were observed
\cite{Schaeffer2017,Schaeffer2019,Yao2021,Tang2025,Johnson2025}.
Such density structures provide important information on the shock-formation process.
However, when the upstream magnetized plasma is spatially non-uniform,
an observed density variation cannot always be uniquely attributed to reflected ions or to the shock front, 
because it may also contain contributions from pre-existing upstream non-uniformity and from the piston--ambient interaction
\cite{Tang2025,Johnson2025}.
This ambiguity is particularly important during the formation stage of supercritical magnetized collisionless shocks,
when a density jump is expected at the shock front and the density can become particularly large
in the overshoot region within the shock transition layer.
At the same time, similarly pronounced density variations may also occur at the electron-scale interface
between the two plasmas when the densities of the two interacting plasmas differ substantially.
Recent experiments have therefore emphasized the importance of separating the shock structure from the driving piston and of following the temporal development of the shock
\cite{Tang2025,Valenzuela-Villaseca2026}.
Therefore, before discussing reflected-ion signatures in the shock transition layer,
the electron-scale interface between the piston and ambient plasmas should first be identified independently.

In this paper, we extend our previous experiments on quasi-perpendicular supercritical magnetized collisionless shocks
\cite{Yamazaki2022,Matsukiyo2022}. 
In those experiments, the ambient-gas pressure was higher than in the present study, specifically by a factor of five, and the resulting brighter self-emission from the ambient plasma made the interface region difficult to identify clearly.
By reducing the ambient-gas pressure, the outer bright front can be identified much more clearly.
A characteristic feature of our experimental system is the almost spatially uniform mass density of the ambient magnetized plasma, 
which results in an approximately uniform Alfv\'en velocity in the region of interest.
This well-controlled upstream condition allows us to identify the interface between the piston plasma and the ambient magnetized plasma unambiguously.
This feature represents a clear advantage of our experimental system,
especially when compared with earlier experiments in which spatial
non-uniformity of the upstream plasma complicated the interpretation of the
shock-formation process.

The present paper and the companion paper \cite{Yamazaki2025a} play complementary roles.
While the companion paper focuses on reflected-ion signatures and the associated  ion temperature anisotropy,  
the present paper uses self-emission imaging to locate the electron-scale interface between the piston and ambient plasmas, thereby enabling the shock structure to be distinguished from the interface itself.
In particular, we demonstrate that the bright outer structure observed in the self-emission images corresponds to the leading edge of the piston plasma, i.e., the electron-scale interface between the piston and ambient plasmas.
This identification enables a clear spatial separation between this interface and the forming shock structure on its upstream side, allowing their roles to be discussed independently.

This paper is organized as follows.
In Sec.~\ref{sec:setup}, we describe the experimental setup and diagnostics.
In Sec.~\ref{sec:analysis}, we present the self-emission imaging results and report the temporal
evolution of the interface structure and its dependence on the magnetic-field
polarity.
In Sec.~\ref{sec:mhdsim}, we compare the experimental observations with two-dimensional
ideal-magnetohydrodynamics (MHD) simulations
and discuss the physical interpretation of the observed structures.
Sec.~\ref{sec:discuss} is devoted to discussion.
Finally in Sec.~\ref{sec:conclusion} we summarize the main conclusions of this study.

\section{Experimental Setup}\label{sec:setup}

The overall experimental setup and basic concept follow those employed in our
previous experiments~\cite{Matsukiyo2022,Yamazaki2022}.
In this study, we focus particularly on self-emission imaging and perform a
more detailed analysis of the effects of external magnetic-field polarity and
the properties of the outer emission front.

\begin{figure}[t]
\includegraphics[width=110mm, bb= 0 0 900 1000]{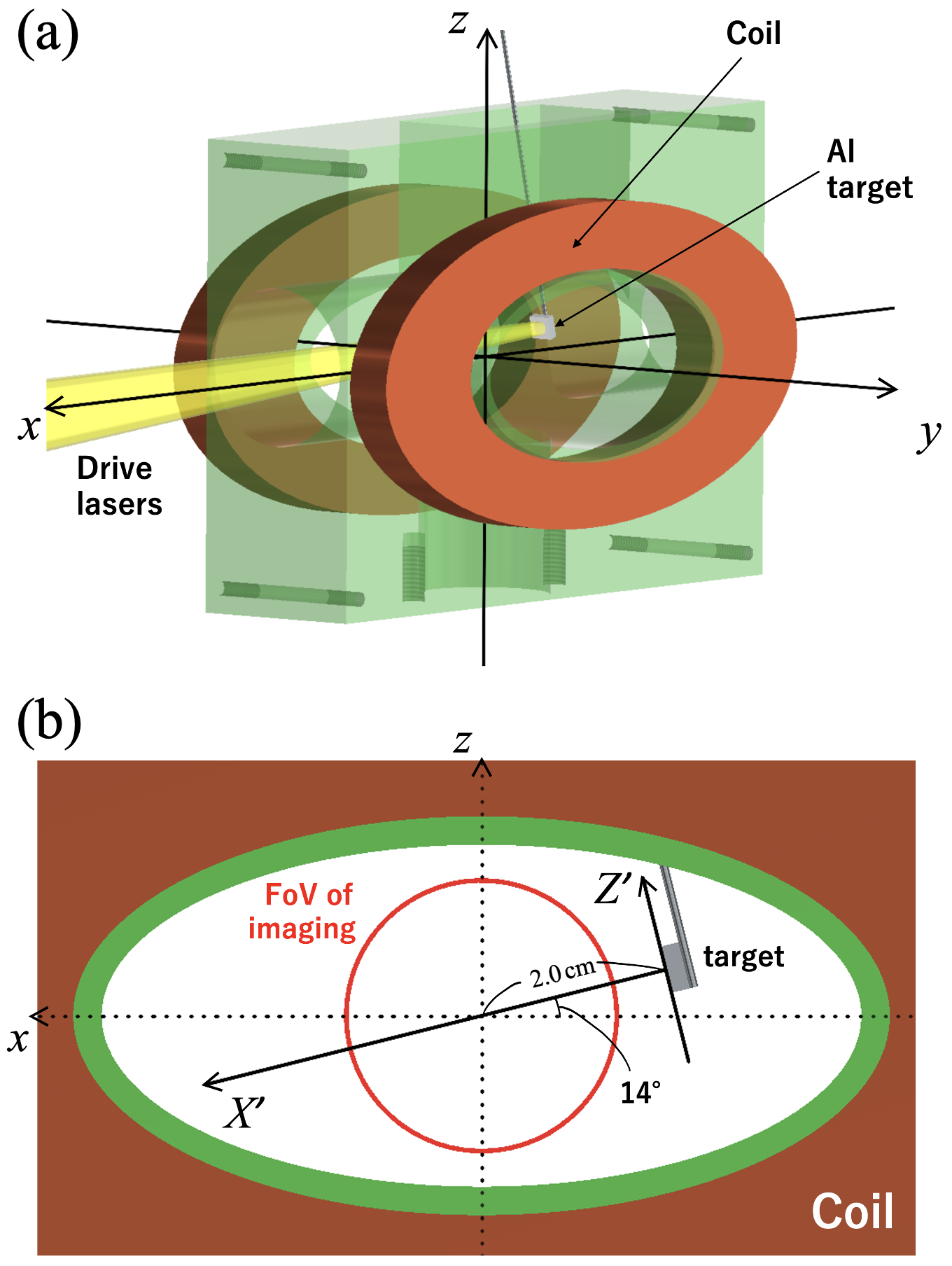}
\caption{
(a) Bird's eye view of experimental setup. Green-colored jig to fix the coil is made by Glass Fiber Reinforced Plastic.
(b)~Side view of our experiment (from $y$-axis).
The red circle, whose center and radius are TCC ($x=y=z=0$) and 1.4~cm,
means field of view of imaging measurement of the plasma self emission.
We define $X'$ axis along the target normal, and $Z'$ axis perpendicular to $X'$ and $y$ axes.
The target position is $X'=0$, and TCC corresponds to $X'=2.0$~cm.
\label{fig:setup}
}
\end{figure}

A Cartesian coordinate system $(x, y, z)$ was defined with its origin at the
target chamber center (TCC) (see Fig.~\ref{fig:setup}).  
The definition is the same as adopted in Yamazaki et al. \cite{Yamazaki2022}.
The $z$ axis was taken to be vertical, and the main drive laser propagated
nearly along the $x$ axis toward the aluminum (Al) target.  
The $y$ axis was then chosen so that the $(x, y, z)$ system
forms a right-handed coordinate system.
The Al target was a $5~\mathrm{mm} \times 5~\mathrm{mm}$ plate with a
thickness of $2~\mathrm{mm}$.  
The target normal lies in the $x$-$z$ plane ($y=0$), and makes an angle of 14$^\circ$ with
$x$-axis (see Fig.~\ref{fig:setup}b).
Unless otherwise noted, the experiments were conducted with an ambient
nitrogen (N) gas pressure of $P_{\rm N} = 1.0~\mathrm{Torr}$ before the laser 
irradiation.  
Just before each shot, the external magnetic field was applied as described
later.  
The target was irradiated by four beams from the Gekko-XII HIPER Laser system,
with wavelength $1053~\mathrm{nm}$, pulse duration $1.3~\mathrm{ns}$,
average energy of $671\pm43$~J~per~beam, and a focal-spot diameter of $2.8~\mathrm{mm}$.
Ionizing photons generated by the laser-produced Al plasma  ionized
the ambient N gas, producing a magnetized N plasma.
Subsequently, expanding Al plasma pushed the magnetized N plasma outward,
thereby creating an interaction region between the two plasmas.  
Compared with the 5-Torr experiments in our previous studies
\cite{Matsukiyo2022,Yamazaki2022}, the lower N pressure of
$P_{\rm N} = 1$~Torr used here reduces collisional effects in the N plasma whose ion density is 
$n_{\rm N}=0.64\times10^{17}$~cm$^{-3}$ when the N gas is fully ionized,
and makes the Al--N plasma interaction more collisionless.  
The self-emission structures analyzed in the following sections originate from
this interaction region.
The geometry of the coil and target is the same as in the companion
paper~\cite{Yamazaki2025a}.

We next describe the external magnetic-field system used in the present experiment.
As shown in Fig.~\ref{fig:setup}(a),
a Helmholtz-type coil was employed to apply a uniform magnetic field in the $y$-direction. 
Since the shock propagates approximately along the target-normal direction in the $x$-$z$ plane, the upstream magnetic field is nearly perpendicular to the shock normal; thus, the shock geometry is quasi-perpendicular.
To observe the long-time evolution of the shock,
the coil was designed with an elliptical shape
so as to elongate the spatial region along the shock-propagation direction.
The coils were fabricated by winding two layers of 50 turns of
flat copper wire ($5 \times 0.5$~mm$^2$),
resulting in a total of 100 turns.
The inner major and minor axes of the elliptical coil
are 87~mm and 43~mm, respectively.
The two coils were placed parallel to and symmetric with respect to the $x$-$z$ plane
with a separation of 39~mm.
The coil is connected to a pulsed-power supply \cite{Edamoto18}
consisting of a capacitor bank (capacitance of 12~mF, voltage of 1.5~kV)
and a switching circuit.
A discharge current (peak of 4.2~kA and pulse width of 5.4~ms FWHM)
flows through the coils to drive the external magnetic field.
The timing of the switching circuit is synchronized with the laser irradiation,
and the $y$-component of the magnetic field around TCC, $B_{y0}$, reaches
$|B_{y0}|\approx4$~T at the current peak.
Along the $y$ axis, we had $B_x=B_z=0$, so that
the external magnetic field is dominated by $y$ component in the region of our interest.
By changing the connection between the coil and the pulsed-power supply
to reverse the direction of the current flowing through the coil,
the direction of the magnetic field is also reversed.

The self-emission images were recorded using a gated optical imager (GOI)
viewing from the $+y$ direction
through a blue filter (central wavelength $450 \pm 2$~nm, FWHM $10 \pm 2$~nm), 
so that the imaging plane is perpendicular
to the $y$ axis.  
The GOI field of view, indicated by the red circle in Fig.~\ref{fig:setup}(b), 
is centered at the TCC and has a radius of 1.4~cm.
As shown in Fig.~\ref{fig:setup}(b), a coordinate system $(X', Z')$ in $y=0$ plane
was defined for the analysis.  
The $X'$ axis is along the target-normal direction,
and the $Z'$ axis is defined to be perpendicular to both the $y$ and $X'$ axis, 
forming an orthogonal coordinate system.
The exposure time of the GOI was $0.2~\mathrm{ns}$.
Note that a single GOI image was acquired at a specified delay time in each shot.

\section{Analysis of Experimental Data}
\label{sec:analysis}

In this section, we show the results of self-emission imaging with GOI.
The sign of the 
$y$-component of  the applied magnetic field near TCC provides a convenient way to characterize the experimental conditions.
In this paper, we symbolically refer to the three representative cases as 
``$B_{y0}=-4$~T",
 ``$B_{y0}=4$~T",
 and ``$B_{y0}=0$",
according to the magnetic-field polarity (or absence) around the TCC.

\begin{figure*}[t]
\includegraphics[width=170mm, bb= 0 0 1150 600]{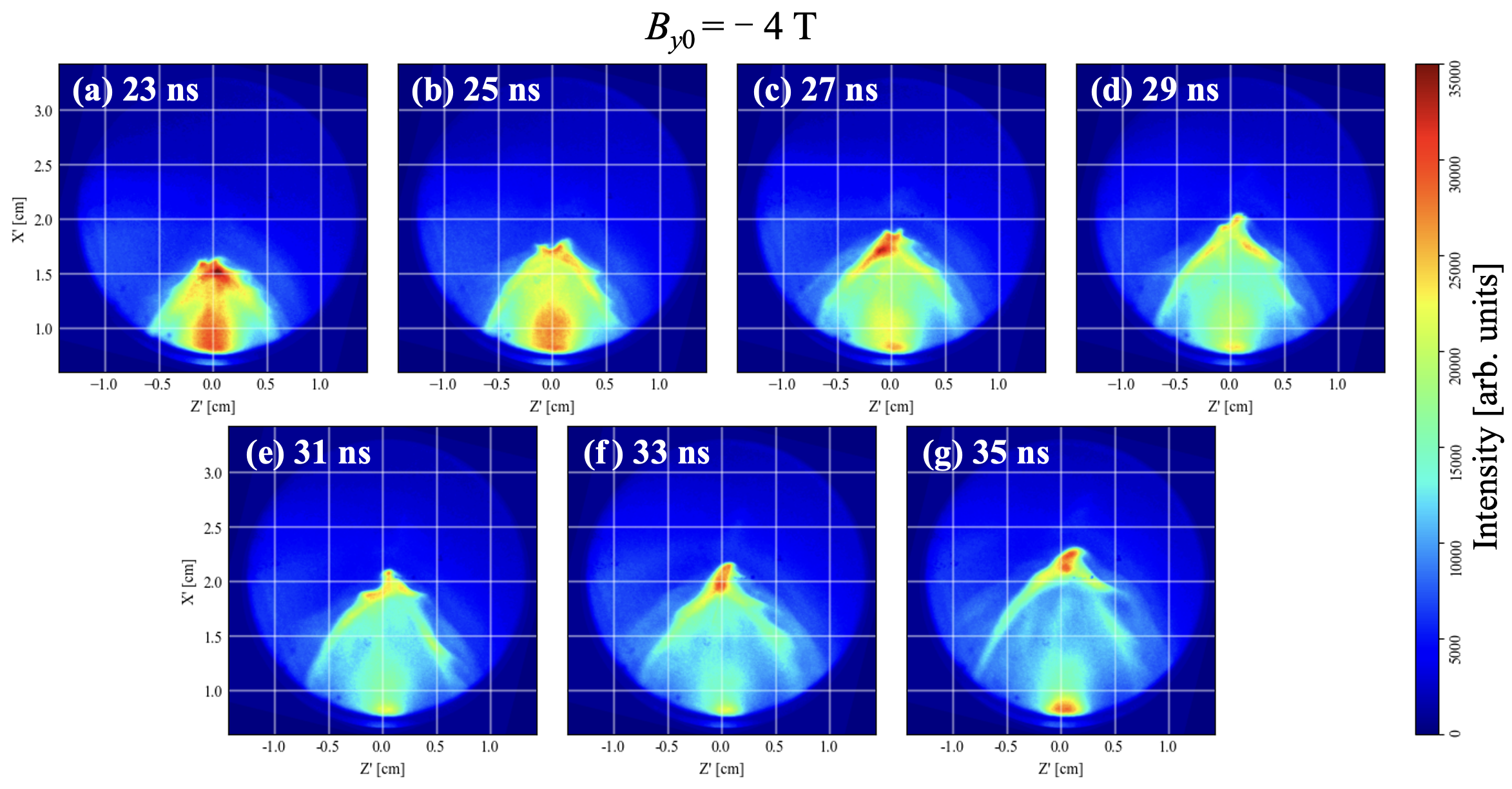}
\includegraphics[width=140mm, bb= 0 0 950 350]{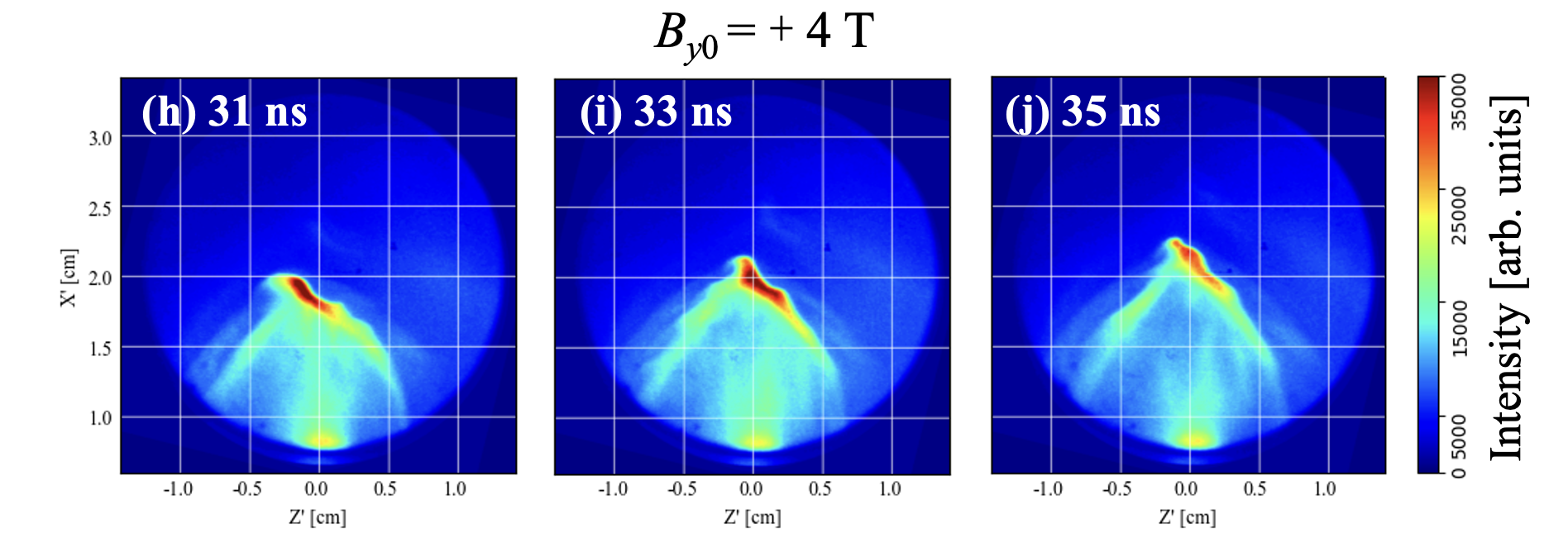}
\caption{
Results of self-emission imaging for different delays after the shot in the presence of the external magnetic field.
Panels~(a)--(g) show the time evolution from 23 to 35~ns for $B_{y0}=-4$~T, while panels~(h)--(j) show the images at 31, 33, and 35~ns for $B_{y0}=+4$~T.
See Fig.~\ref{fig:setup}(b) for the definitions of $X'$ and $Z'$ axes.
\label{fig:SE_By-}
}
\end{figure*}

Figure~\ref{fig:SE_By-} shows the time evolution of the self-emission images for both magnetic-field polarities. 
For $B_{y0}=-4~\mathrm{T}$,
as time increases from 23~ns [panel~(a)] to 35~ns [panel~(g)], the bright region expands outward.
The apparent propagation velocity of the front is 
$V_{X'}\approx500$~km~s$^{-1}$.
A similar outward expansion is also observed for $B_{y0}=+4~\mathrm{T}$ 
[panels~(h)--(j)], with a propagation velocity comparable to that for $B_{y0}=-4~\mathrm{T}$.
Since the Alfv\'{e}n speed of upstream plasma with $|B_{y0}| = 4~\mathrm{T}$ is estimated as
$v_{\rm A}=|B_{y0}|/\sqrt{4\pi m_{\rm N}n_{\rm N}}\approx92$~km~s$^{-1}$, where $m_{\rm N}$ is the mass of N ion,
it is expected that the generated shock in the magnetized N plasma is supercritical.

It is found from Fig.~\ref{fig:SE_By-} that
the outer boundary of the brightest area is not perfectly smooth 
but exhibits small-scale ripples that become more noticeable at later times.
Such ripple structures are visible for both magnetic-field polarities.
For $B_{y0}=-4~\mathrm{T}$,
the overall emission pattern is asymmetric with respect to the $X'$ axis.
Along the propagation direction (the target normal, $X'$ axis), 
the ripples develop mainly on the right-hand side, 
while those on the left are less pronounced.
The self-emission intensity is brighter on the left side 
than on the right.
Thus, the ripples are more pronounced on the weak-emission side.
Note that after 27~ns, a bow-shock-like dim emission can be recognized upstream of the brightest regions 
[see Fig.~\ref{fig:SE_By-}(f) for example].
This structure is further discussed in the companion paper~\cite{Yamazaki2025a}.

\begin{figure*}[t]
\includegraphics[width=170mm, bb= 0 0 1150 380]{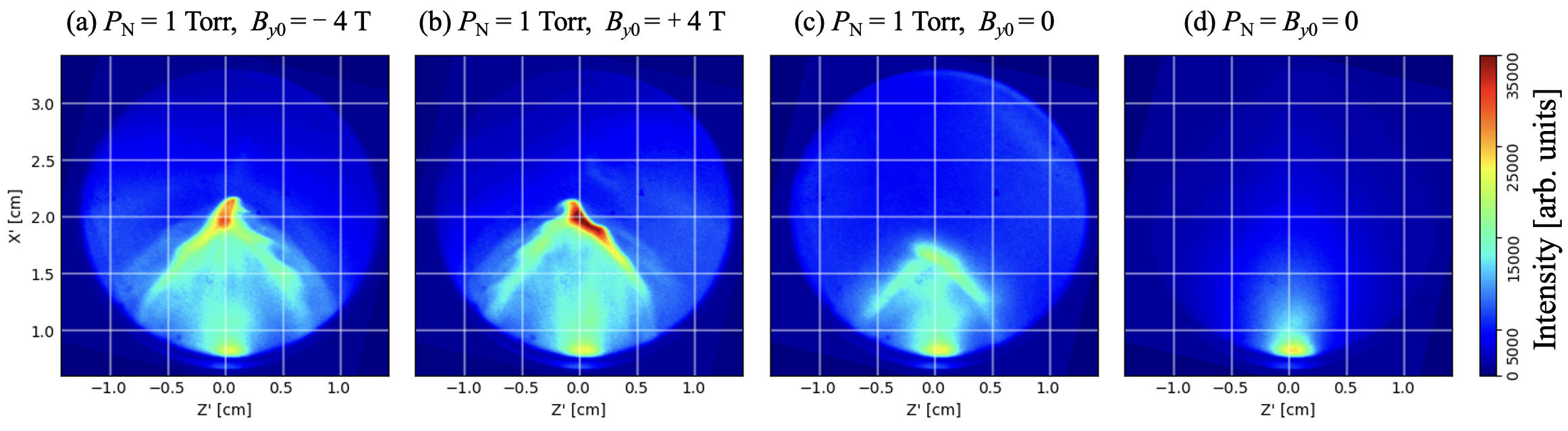}
\caption{
Results of self-emission imaging at 33~ns after the shot
for 
(a)~$P_{\rm N}=1$~Torr and  $B_{y0}=-4$~T,
(b)~$P_{\rm N}=1$~Torr and  $B_{y0}=+4$~T,
(c)~$P_{\rm N}=1$~Torr and  $B_{y0}=0$,
and (d)~$P_{\rm N}=B_{y0}=0$.
See Fig.~\ref{fig:setup}(b) for the definitions of $X'$ and $Z'$ axes.
\label{fig:SE_33ns}
}
\end{figure*}

Figure~\ref{fig:SE_33ns} shows the self-emission images at 33~ns 
for the four  cases: 
(a)~$B_{y0} = -4~\mathrm{T}$, (b)~$B_{y0} =+4~\mathrm{T}$, and (c)~$B_{y0} =0~\mathrm{T}$ 
with ambient N gas before the shot ($P_{\rm N}=1$~Torr),
and (d)~the shot with neither the ambient N gas nor external magnetic field ($P_{\rm N}= B_{y0} =0$).
The images for $B_{y0} = -4~\mathrm{T}$ and $+4~\mathrm{T}$ 
are nearly mirror-symmetric to each other with respect to the $X'$ axis, 
indicating that the overall emission pattern reverses 
when the magnetic-field polarity is changed.
For the case without the external magnetic field ($B_{y0} = 0$) but with ambient N gas $P_{\rm N}=1$~Torr, 
the overall emission intensity is weaker, and the outer boundary of the bright region 
appears less distinct.  
Moreover,  the outer boundary of the bright area seems to remain slightly closer to the target 
compared with the cases with the magnetic field.

\begin{figure}[t]
\includegraphics[width=90mm, bb= 0 0 1000 700]{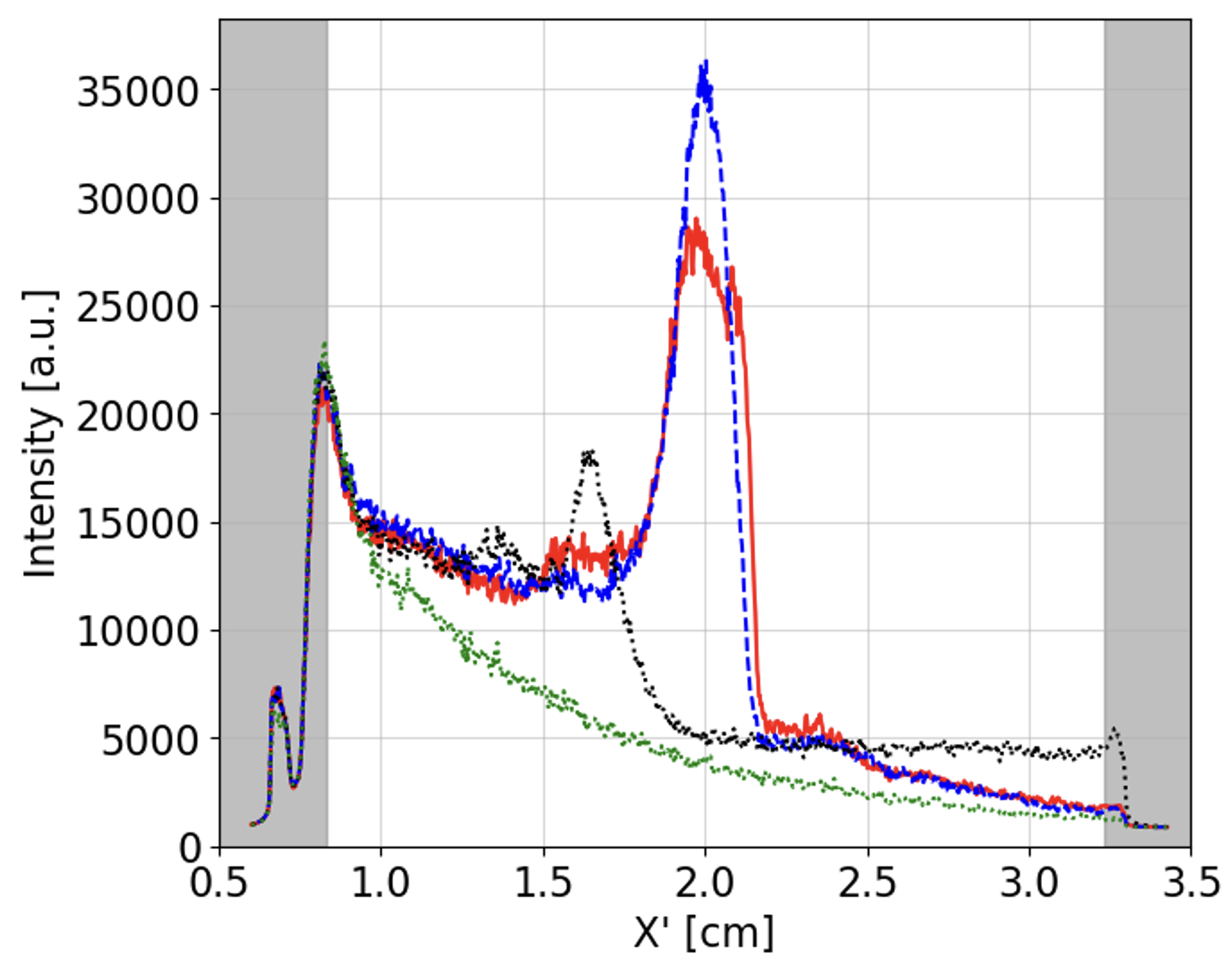}
\caption{
Spatial distributions of self-emission along $X'$-axis ($Z'=0$) at 33~ns after the shot for
the cases of $B_{y0}=-4$~T (red solid line), $B_{y0}=4$~T (blue dashed line), $B_{y0}=0$ and $P_{\rm N}=1$~Torr (black dotted line),
and $B_{y0}=P_{\rm N}=0$ (green dotted line).
See Fig.~\ref{fig:setup}(b) for the definitions of $X'$ and $Z'$ axes.
In gray-shaded regions, the detector gain is small at the boundary of the field of view. 
\label{fig:SE_Xslice}
}
\end{figure}

Figure~\ref{fig:SE_Xslice} 
shows the spatial distributions of self-emission along the $X'$ axis ($Z' = 0$) at 33~ns after the shot for 
the cases of $B_{y0} = -4~\mathrm{T}$ (red solid line), $B_{y0} = +4~\mathrm{T}$ (blue dashed line), and $B_{y0} = 0$ with $P_{\rm N} = 1~\mathrm{Torr}$ (black dotted line), 
while the green dotted line represents the case of $B_{y0} = P_{\rm N} = 0$.  
All the profiles for the cases with $P_{\rm N} = 1~\mathrm{Torr}$ exhibit a distinct intensity maximum resulting from the interaction between the Al and N plasmas, 
and the peak positions correspond to the locations of the outer boundaries of the bright regions shown in Fig.~\ref{fig:SE_33ns}.  
When the external magnetic field is applied (red and blue lines), the peaks appear at $X' \approx 2.0~\mathrm{cm}$, 
and the profiles show a sharp drop at $X' \approx 2.1$--2.2~cm, followed by gradual emission tails in the region $X' > 2.5~\mathrm{cm}$.  
Without the external magnetic field (black dotted line, $P_{\rm N} = 1~\mathrm{Torr}$), the peak is located at $X' \approx 1.6~\mathrm{cm}$, 
which is slightly inside compared with the cases of $B_{y0} = \pm 4~\mathrm{T}$.  
This profile shows a less steep decay with increasing $X'$ than those with the external magnetic field, 
and the emission intensity remains almost flat for $X' > 2.0~\mathrm{cm}$.  
In the inner region ($X' < 1.0~\mathrm{cm}$), excluding the gray-shaded area at the edge of the field of view, 
all the profiles nearly overlap with each other, 
indicating that this inner region is clearly dominated by the Al plasma; the Al-dominated region extends farther outward toward the bright self-emission peak associated with the Al--N interface, as discussed in Sec.~\ref{subsec:iondynamics}.

\begin{figure}[t]
\includegraphics[width=90mm, bb= 0 0 750 580]{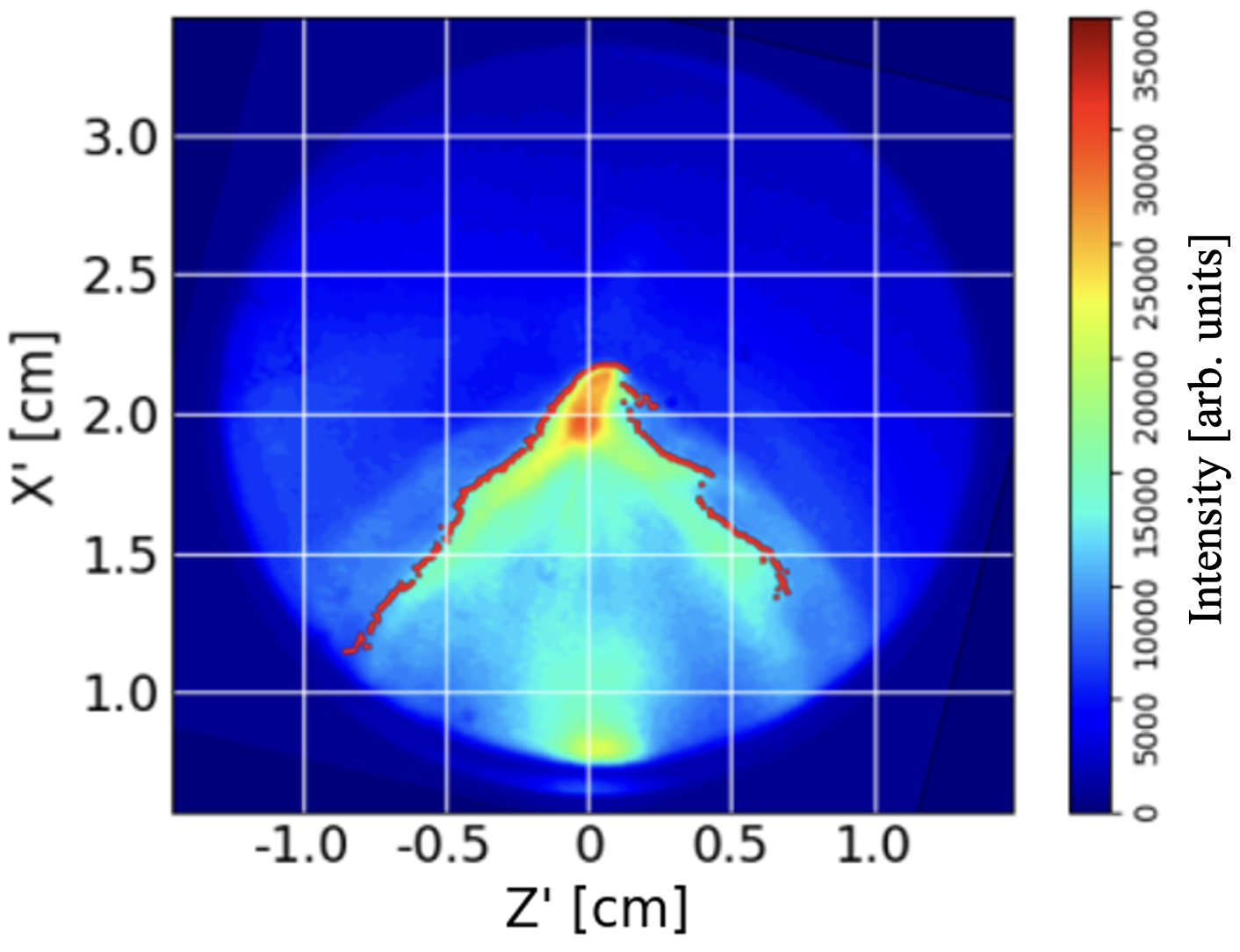}
\caption{
Self-emission image at 33~ns after the shot for $B_{y0}=-4$~T, which is overlaid with red points 
representing the local maximum of the rate of change, $|\vec{\nabla} I|^2/I^2$
(see text for details).
Our analysis captures the region at which the emission steeply changes.
\label{fig:SE_grad_33ns_By-}
}
\end{figure}

When the external magnetic field is applied, as shown in Figs.~\ref{fig:SE_33ns} and \ref{fig:SE_Xslice}, 
the outer boundary of the emission region becomes sharper.  
To determine the boundary position quantitatively, we analyze the self-emission
intensity $I(X',Z')$ by evaluating the normalized squared gradient, $|\vec{\nabla} I|^2/I^2$, where
$\vec{\nabla}=(\partial/\partial X',\,\partial/\partial Z')$ is the two-dimensional gradient operator.
Since the images are defined on discrete pixels,
the spatial derivatives are evaluated using finite differences.
We denote the intensity at the pixel in the $i$th row and $j$th column as $I_{ij}$, and calculate
\begin{equation}
r_{ij} =
\frac{(I_{i+1,j}-I_{i-1,j})^2+(I_{i,j+1}-I_{i,j-1})^2}{I_{ij}^2}~~,
\label{eq:gradient}
\end{equation}
as a discrete analogue of $|\vec{\nabla} I|^2/I^2$.
The local maxima of $r_{ij}$ correspond to locations where the self-emission intensity changes rapidly,
thereby making the ripple structures shown in Figs.~\ref{fig:SE_By-} and \ref{fig:SE_33ns} easier to visualize.
Figure~\ref{fig:SE_grad_33ns_By-} shows the original emission distribution $I_{ij}$, identical to
Fig.~\ref{fig:SE_By-}(f) and Fig.~\ref{fig:SE_33ns}(a) ($B_{y0} = -4~\mathrm{T}$ at $t = 33$~ns),
overlaid with the local maxima of $r_{ij}$ (red dots), which correspond to the extracted boundary positions.
These red dots are aligned along the locations where the emission intensity decreases sharply outward,
indicating that the maxima of the normalized rate were correctly extracted.
This result demonstrates that the present analysis method is effective for quantitatively identifying
the shape and spatiotemporal evolution of the outer boundary.

\begin{figure}[t]
\includegraphics[width=105mm, bb= 0 0 1300 580]{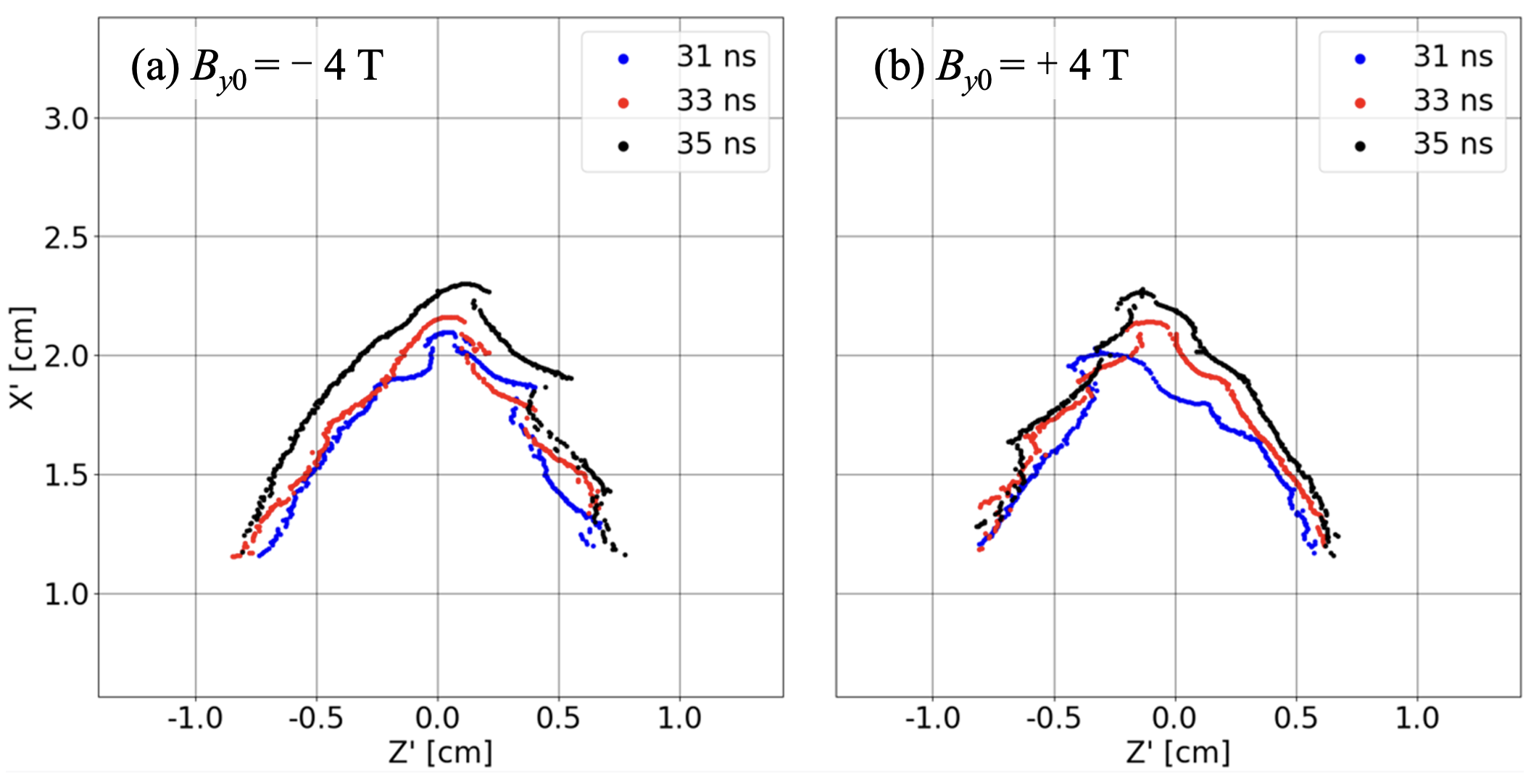}
\caption{
Red points of left and right panels show the 
position of local maximum of $|\vec{\nabla} I|^2/I^2$, 33~ns after the shot
for $B_{y0}=-4$ and 4~T, respectively.
Red points in the left panel are identical to those shown in Fig.~\ref{fig:SE_grad_33ns_By-}.
Blue and black points are the same as red points but for 31 and 35~ns after the shot,
respectively.
\label{fig:SE_grad_31_33_35ns}
}
\end{figure}

Figure~\ref{fig:SE_grad_31_33_35ns} shows the positions of the local maxima of the normalized intensity change rate, 
$|\vec{\nabla} I|^2 / I^2$, extracted in the same manner as in Fig.~\ref{fig:SE_grad_33ns_By-}, 
for (a) $B_{y0} = -4$~T (left panel) and (b) $B_{y0}=+4$~T (right panel) at $t = 31$, 33, and 35~ns.  
The red dots in Fig.~\ref{fig:SE_grad_31_33_35ns}(a) are identical to those plotted in Fig.~\ref{fig:SE_grad_33ns_By-}.
The results for $B_{y0} = -4~\mathrm{T}$ and $+4~\mathrm{T}$ are nearly mirror-symmetric 
with respect to the $X'$ axis, 
quantitatively confirming the reversed asymmetry of the self-emission images shown in Fig.~\ref{fig:SE_33ns}(a) and (b).  
As time progresses, the outer boundaries move outward, 
corresponding to the expansion of the bright region observed in Fig.~\ref{fig:SE_By-}.
In addition, small-scale ripple-like modulations are seen along the boundary, 
and their amplitude and spacing exhibit slight variations with time.  
Although the shape of ripples appears to change gradually with  time,
it was difficult to quantify their wavelength or amplitude, 
because typically only one  ripple wavelength is observed.  
Nevertheless, this qualitative trend is consistent with the growth of small-scale structures observed 
in the late-time images of Fig.~\ref{fig:SE_By-}.

\section{MHD simulations}
\label{sec:mhdsim}

In the previous section, we analyzed the spatiotemporal evolution of the
self-emission structures, and demonstrated that the outer emission front and
its ripple-like modulations exhibit a clear dependence on the polarity of the
applied magnetic field.  
To interpret these features and to clarify the underlying plasma dynamics,
we compare the experimental observations with two-dimensional MHD simulations.

\subsection{Validity of fluid approximation around the interface between two plasmas}
\label{subsec:fluidapprox}

First, we evaluate the validity of the MHD approximation adopted here.
To determine the plasma parameters in the region where the bright self-emission is observed,
we  performed the collective Thomson scattering (CTS) measurements
(for details, see \S~\ref{sec:TSmeasure_SOP}).
From these measurements, we obtained for the Al plasma
an electron density $n_e \approx2 \times 10^{18}~\mathrm{cm^{-3}}$,
an ion charge state $Z = 11$,
and an ion density $n_i =n_e/Z\approx2 \times 10^{17}~\mathrm{cm^{-3}}$.
Using these parameters, we calculate the ion inertial length
$c/\omega_{pi} \approx 0.024$~cm.
In addition, if we adopt the magnetic-field strength, $B=10$~T, and ion velocity, $v_i=500$~km~s$^{-1}$,
around the TCC at $t \approx 33$~ns,
then we get the ion gyro-radius
$r_{g,i} = m_{\rm Al} c v_0 /Z e B \approx 0.13$~cm
and  the ion--ion Coulomb mean free path $\lambda_{ii} \approx 0.47$~cm.
Since the ion inertial length is sufficiently smaller than the 
characteristic scale length of the measured ripples ($\approx0.1$--0.4~cm) and
the size of the rounded leading edge with bright self-emission ($\approx0.2$--0.4~cm),
the fluid approximation is justified for the high-density, bright-emission region
around the interface between N and Al plasmas.
Note that, for the same representative parameters, the Al-ion gyro-period is approximately 16~ns, which is slightly smaller than the characteristic evolution timescale of the experiment. This indicates that Al-ion gyromotion may contribute to the nonstationary evolution of the interface. Nevertheless, the present MHD simulations are intended to describe qualitatively the global morphology of the high-density Al--N interface rather than its ion-scale kinetic dynamics.

\subsection{Simulation setup}

The simulations solve the two-dimensional ideal MHD equations and were performed using the open-source code, 
Athena++ \cite{White2016,Stone2020}, adopting Cartesian coordinates. 
The calculation is based on single-fluid MHD; the Al and N plasmas are not treated as separate fluids but are represented by different initial and boundary conditions.
We employ Harten-Lax-van Leer-Einfeldt (HLLE) Riemann solver, a second-order piecewise linear reconstruction, and a second-order Runge-Kutta time integrator.

\begin{figure}[t]
\includegraphics[width=140mm, bb= 0 0 1500 900]{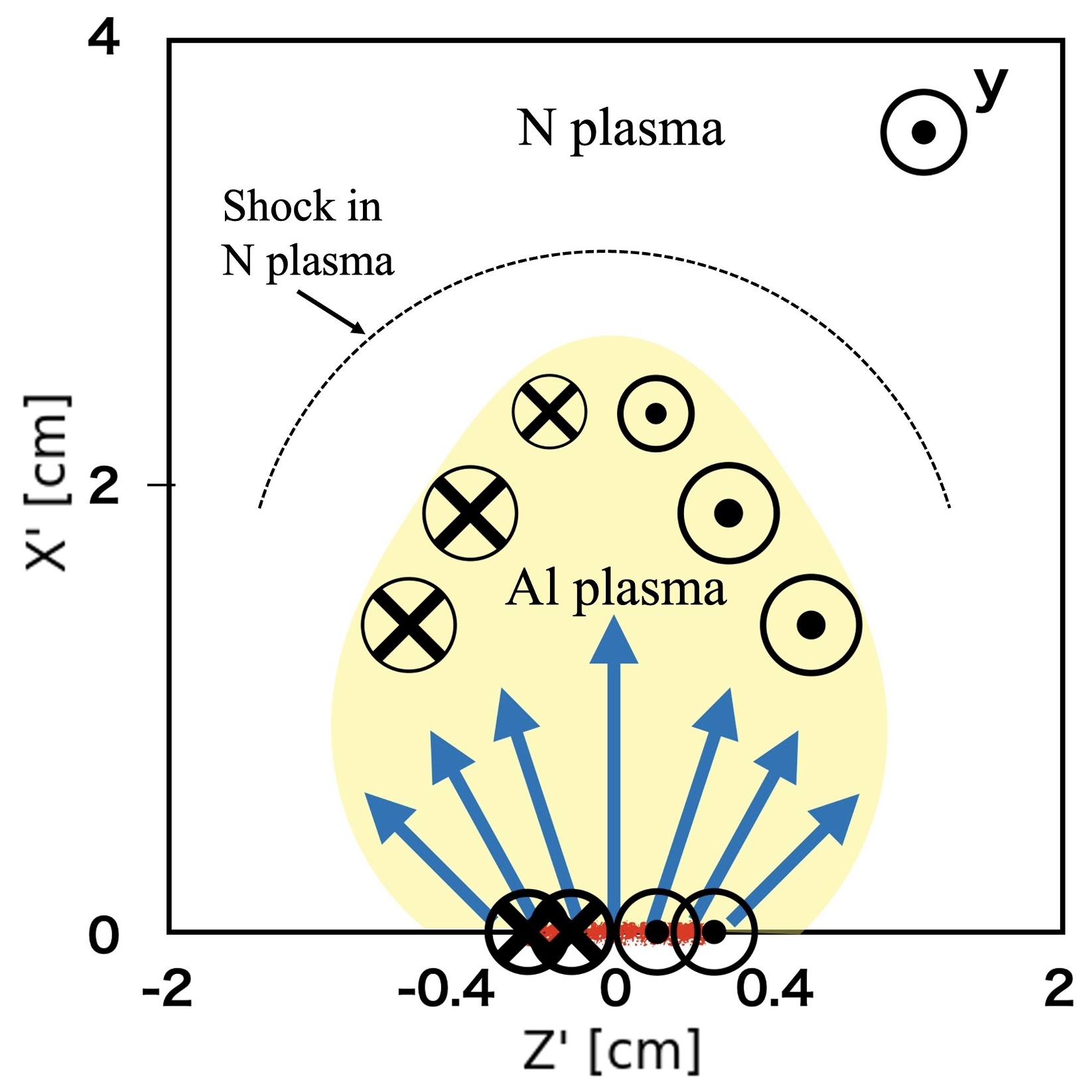}
\caption{
Schematic illustration of the simulation setup in the $X'$-$Z'$ plane. 
Al plasma (yellow region) is injected from the boundary (the red line on the $Z'$ axis, $X'=0$) with velocity vectors indicated by blue arrows. 
The symbols $\odot$ and $\otimes$ in the Al plasma region denote the out-of-plane magnetic field components, $\pm B_y$, to mimic the azimuthal field $B_{\phi}$. 
The setup includes a self-magnetized ejecta expanding into a magnetized ambient N plasma 
(uniform magnetic field $B_{\rm N,0}\hat{\bm e}_y$, 
where $B_{{\rm N},0}$ is constant).
The dashed line represents a shock in the magnetized N plasma.
For $B_{{\rm N},0}>0$ ($B_{{\rm N},0}<0$), the density of the injected Al plasma,
$\rho_{\rm Al}(X'=0,Z')$, is higher (lower) in the region $0<Z'<R_{\rm in}$
than in the region $-R_{\rm in}<Z'<0$,
where $R_{\rm in}=0.4$~cm [see Eqs.~(\ref{eq:rho_B0_posi}) and (\ref{eq:rho_B0_nega})].
\label{fig:MHD_setup}
}
\end{figure}

The computational domain is in the $X'$-$Z'$ plane that is initially filled with an ambient N plasma
with uniform mass density $\rho_{{\rm N},0}$.
As shown in Fig.~\ref{fig:MHD_setup},
the simulation box size is $L_{X'} \times L_{Z'}=4~{\rm cm}\times4~{\rm cm}$ (that is, $0\leq X'\leq4$~cm and $-2\leq Z'\leq2$~cm).
The grid sizes and time steps of the simulations are $\Delta X' =\Delta Z'= 0.002~{\rm cm}$ and $\Delta t=0.05\Delta X'/|V_{\rm Al}|$, where $V_{\rm Al}$ is the velocity of the expanding Al plasma.
The Al plasma is injected at $X'=0$ over the range $Z'\in[-R_{\rm in},R_{\rm in}]$, where $R_{\rm in}=0.4$~cm.
We adopt outflow boundary conditions for all boundaries except for the injection region.
The external uniform magnetic field is imposed perpendicular to the simulation plane, $B_{\rm N,0}\hat{\bm e}_y$.
Here we focus on how the result changes depending on the sign of $B_{\rm N,0}$.

Near the target, expanding Al plasma has an electron density gradient, $\vec{\nabla} n_{\rm e}$, in the $X'$-direction
and the electron pressure gradient, $\vec{\nabla} P_{\rm e}$, toward the center of the laser focal-spot ($X'\approx Z'\approx0$).
Thus, the Biermann battery effect 
($\partial \vec{B}/\partial t \propto \vec{\nabla} P_{\rm e}\times\vec{\nabla} n_{\rm e}$) 
\cite{Biermann1950}
generates toroidal magnetic fields around the $X'$-axis
\cite{Stamper1991,Gregori2012,Kugland2013,Ryutov2013,Umeda2019,Griff-McMahon2024}.
In the present two-dimensional MHD simulations, we adopt a simplified out-of-plane magnetic-field configuration that mimics such a toroidal field, with $B_y>0$ for $Z'>0$ and $B_y<0$ for $Z'<0$, as illustrated in Fig.~\ref{fig:MHD_setup}.

Since our simulations solve neither the radiation transport equation nor the self-generation of magnetic fields due to the 
Biermann battery effect, we continue to inject the magnetized Al plasma with an expanding velocity at the injection boundary for 2.8 ns.
At the injection boundary $X'=0$, we set the magnetic field in the Al plasma $\vec{B}_{\rm Al}(Z')$ perpendicular to the $X'$-$Z'$ plane (along $\pm y$-direction) to approximate the toroidal configuration:
\begin{equation}
\vec{B}_{\rm Al}(X'=0,Z')=B_{\rm Al,0}\tanh(Z'/0.1R_{\rm in}) \hat{\bm e}_y,
\label{eq:B-Biermann}
\end{equation}
where $B_{\rm Al,0}>0$ is a representative value of the field strength in the Al plasma near the target.
To reproduce asymmetry in the results of self-emission imaging [Fig.~\ref{fig:SE_33ns}(a) and (b)],
the mass density $\rho_{\rm Al}$ of the injected Al plasma at the injection boundary ($X'=0$) is assumed to be
asymmetric with respect to the $X'$-axis (see \S~\ref{subsec:densityasym}).
For the case of imposed magnetic field in the N plasma in the $+y$ direction (that is, $B_{\rm N,0}>0$),
we set
\begin{equation}
\rho_{\rm Al}(X'=0, Z') = \rho_{\rm Al,0} + \delta \rho_{\rm Al} \sin(\pi Z'/R_{\rm in}) ,
\label{eq:rho_B0_posi}
\end{equation}
where $\rho_{\rm Al,0}>0$ and $\delta\rho_{\rm Al}>0$ represent the mean value and fluctuation amplitude, respectively, while
\begin{equation}
\rho_{\rm Al}(X'=0, Z') = \rho_{\rm Al,0} - \delta \rho_{\rm Al} \sin(\pi Z'/R_{\rm in}) ,
\label{eq:rho_B0_nega}
\end{equation}
in the case of $B_{\rm N,0}<0$
(for further details, see \S~\ref{subsec:densityasym}).
Finally, the bulk velocity $\vec{V}_{\rm Al}$ of the injected Al plasma is given as
\begin{equation}
\vec{V}_{\rm Al}(X'=0, Z') = \begin{pmatrix} 
v_{\rm Al} \left| \cos \left[ \pi \tan^{-1} \left( \frac{Z'}{2R_{\rm in}} \right) \right] \right| \\
v_{\rm Al} \sin \left[ \pi \tan^{-1} \left( \frac{Z'}{7R_{\rm in}} \right) \right]
\end{pmatrix},
\label{eq:vel_boundary}
\end{equation}
where $v_{\rm Al} > 0$ is taken to be 100 times the sound speed in the N plasma, 
which corresponds to the typical initial ejecta speed of approximately 700~km~s$^{-1}$ observed in the experiments.

To induce instabilities, 
we introduce a white noise with an amplitude of
$\delta \rho_{\rm N}/\rho_{\rm N,0} \in[-0.002,+0.002]$ into the ambient plasma density.
Two runs are presented for the cases of 
$B_{\rm N,0}>0$ and $B_{\rm N,0}<0$,
corresponding to the experiment with $B_{y0}=+4$ and $-4$~T, respectively. 
Table~\ref{tab:table1} summarizes the parameters employed in the simulation.

The present simulations are two-dimensional and do not include fully three-dimensional effects.
As shown in the next section, however, the present two-dimensional model already reproduces the main experimental features relevant to this study, so there is no strong motivation at present to extend the calculation to three dimensions.
In a fully three-dimensional system, magnetic reconnection could occur where two plasmas carrying oppositely directed magnetic fields come into contact.
However, our primary interest here is the Al--N plasma interface, which is located in a high-$\beta$ region downstream of the shock, where the thermal pressure dominates over the magnetic pressure.
Thus, even if magnetic reconnection occurs locally in three dimensions, it is unlikely to substantially modify the overall interface dynamics discussed in the present work.

\begin{table}
  \caption{Plasma parameters in the MHD simulations.
  Quantities $\beta_{\rm Al,0} = 2P_{\rm Al,0}/B_{\rm Al,0}^2$ and 
  $\beta_{\rm N,0} = 2P_{\rm N,0}/B_{\rm N,0}^2$ are respectively plasma beta for Al and N plasma in the code unit, 
  where $p_{\rm Al,0}$ and $p_{\rm N,0}$ are the initial gas pressures of Al and N plasmas, respectively. 
  }
  \label{tab:table1}
  \centering
  \begin{ruledtabular}
  \begin{tabular*}{\linewidth}{@{\extracolsep{\fill}}cccccc@{}}
    $\rho_{\rm Al,0}/\rho_{\rm N,0}$ &
    $T_{\rm Al,0}/T_{\rm N,0}$ &
    $|B_{\rm Al,0}|/|B_{\rm N,0}|$ &
    $\beta_{\rm Al,0}$ &
    $\beta_{\rm N,0}$ &
    $\delta \rho_{\rm Al}/\rho_{\rm Al,0}$ \\
    \hline
    21 & 32 & 5.3 & 1 & 0.08 & 1.4 \\
  \end{tabular*}
  \end{ruledtabular}
\end{table}

\subsection{Simulation Results}

\begin{figure}[t]
\includegraphics[width=125mm, bb= 0 0 1000 900]{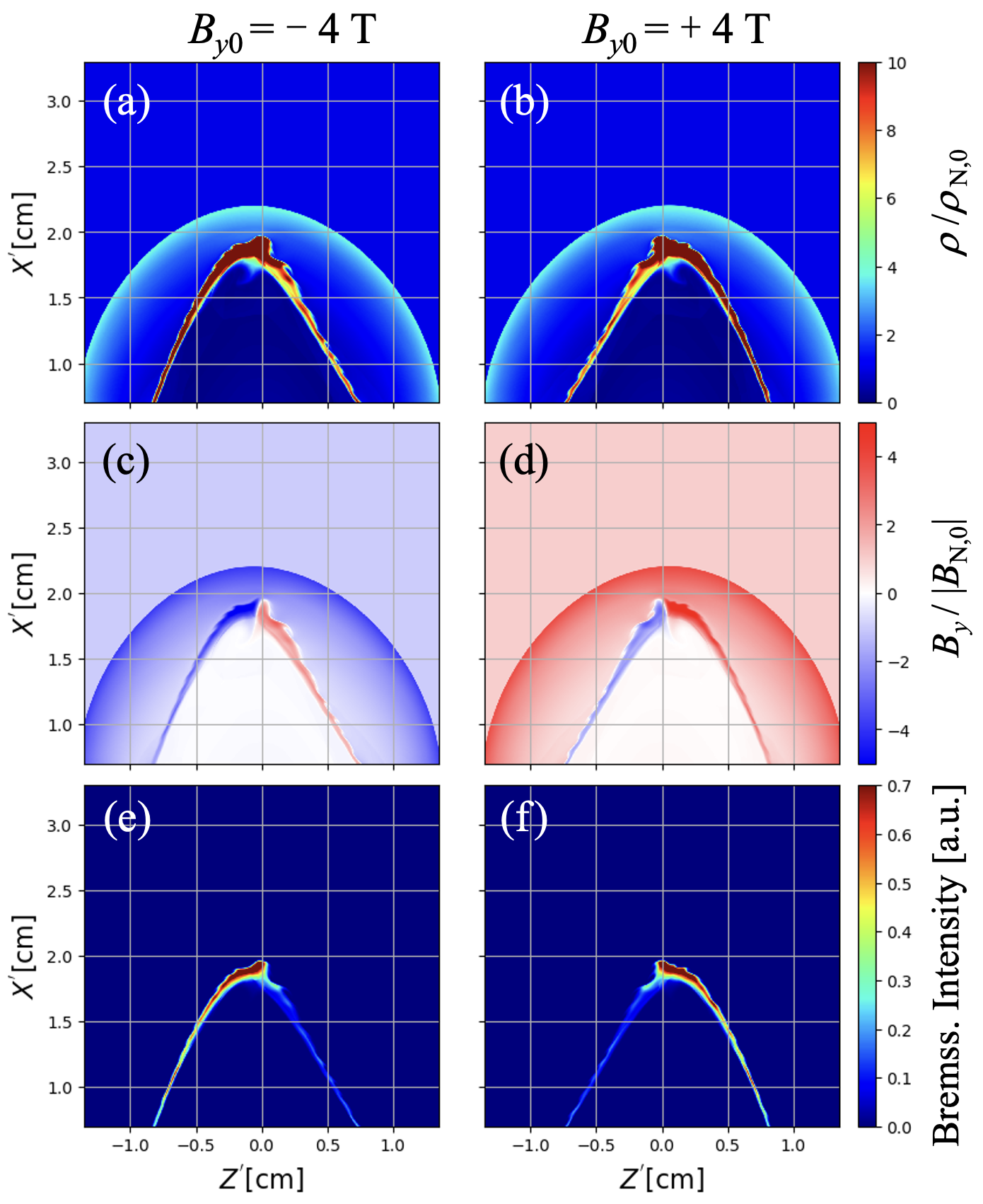}
\caption{
Results of two-dimensional MHD simulations.
All panels are the snap shot  33~ns after the shot.
Left (a, c, and e) and right (b, d, and f) panels are for $B_{y0}= -4$~T and $B_{y0}= +4$~T, respectively.
Top panels (a and b) are for the mass density normalized by the initial density of the N plasma, $\rho_{{\rm N,0}}$.
The middle row (c and d) represents the $y$ component of the magnetic field normalized 
by the  strength initially applied in the N plasma, $|B_{\rm N,0}|$.
Bottom panels (e and f) are the synthetic bremsstrahlung-intensity maps (see Eq.~(\ref{eq:bremss_MHD})).
\label{fig:MHD_33ns}
}
\end{figure}

Figure~\ref{fig:MHD_33ns} shows the MHD simulation results at 33~ns for the two cases of $B_{y0} = -4~\mathrm{T}$ (left panels) and $B_{y0} = +4~\mathrm{T}$ (right panels).  
The density distributions (top row) exhibit a bow-shaped interaction region formed as the expanding Al plasma pushes the magnetized N plasma.
The location and curvature of this high-density layer agree well with the outer emission fronts identified in the GOI images (Figs.~\ref{fig:SE_By-}--\ref{fig:SE_Xslice}).

A clear reversal is observed when the magnetic-field polarity is flipped:
the sign of $B_{y0}$ across the interaction region (middle row) changes,
and the density ridge tilts in the opposite direction 
 (see \S~\ref{subsec:densityasym} for details).
This left-right reversal mirrors the polarity-dependent inversion observed in the self-emission images (Fig.~\ref{fig:SE_33ns}), indicating that the observed asymmetry originates from the global magnetic-field configuration.

We compare the self-emission imaging from the MHD simulation  with the experimental results.
Using the mass density $\rho_s(X',Z')$ and pressure $P_s(X',Z')$, obtained from the MHD simulations, where $s={\rm Al},~{\rm N}$ denotes the plasma species, 
we calculate the synthetic bremsstrahlung intensity at frequency $\nu$ as
\begin{eqnarray}   
    I_s(X', Z') &\propto & n_e n_s Z_s^2 T_s^{-1/2}  e^{-h\nu/k_BT_s} \nonumber \\
    &\propto & \sqrt{\frac{\rho_s(X', Z')^5}{P_s(X', Z')}} \frac{Z_s^3}{A_s^{5/2}}
    \exp\left[ -\frac{h\nu\rho_s(X',Z')}{P_s(X',Z')A_s}\right], \nonumber\\
    &&
\label{eq:bremss_MHD}
\end{eqnarray}
where $Z_s$ and $A_s$ are the ion charge number and atomic mass number, respectively,
and $h$ and $k_B$ are the Planck constant and the Boltzmann constant, respectively.
Here we adopt $Z_{\rm Al}=9$, $Z_{\rm N}=3$, $A_{\rm Al}=27$, and $A_{\rm N}=14$.
The synthetic bremsstrahlung-intensity maps [Fig.~\ref{fig:MHD_33ns}(e) and (f)] highlight the
compressed region at the plasma-plasma interface and reproduce the qualitative brightness distribution seen in the experiment.
In particular, the simulation captures the asymmetric enhancement on the
left or right side depending on the sign of $B_{y0}$, consistent with the measured self-emission profiles and the boundary locations extracted in Fig.~\ref{fig:SE_grad_31_33_35ns}.
We also find that the interface is clearly rippled where the magnetic fields are anti-parallel and the density is lower, as seen in Fig.~\ref{fig:SE_33ns}.

We note that the experimental self-emission images are integrated along the line of sight, whereas the present two-dimensional simulations do not include this projection effect.
The line-of-sight integration can affect the absolute emission intensity and may broaden spatial structures.
However, the asymmetry and ripple structures discussed here are observed near the outer edge of the ejecta, where line-of-sight overlap is geometrically reduced.
We therefore expect that the observed relative asymmetry is only weakly affected by the projection effect.

A straightforward interpretation of the ripple formation might be attributed to
the development of the Kelvin-Helmholtz (KH) instability at the interface between
Al and N plasmas.
The KH instability grows when the sonic Mach number $M_s$ around the interface is less than about unity \cite{Miura1982}.
To examine this possibility, we analyzed the simulation data and evaluated  $M_s$ around the interface.
At early times, $M_s$ is typically $\sim 0.1$, while at a late time
($t = 33$~ns; Fig.~\ref{fig:MHD_33ns}) it becomes of order unity along the interface.
This indicates that the KH mode in the present configuration is marginally unstable,
rather than strongly unstable.
Furthermore, assuming a finite-width velocity shear layer, we estimate the growth
 time of the KH instability for a wavelength of $\sim 0.15$~cm to be $\sim 100$~ns, 
 which is significantly
longer than the timescale of the system.
Therefore, the KH instability is unlikely to be the primary mechanism responsible
for the observed ripples.

We instead consider an alternative interpretation in which the ripple structure
is associated with vortical motions generated around the high-density, rounded leading edge
of the bright emission region near the $X'$ axis.
The formation of this rounded leading edge can be understood from the magnetic-field
and flow structures of the injected Al plasma.
At the injection boundary ($X'=0$), the $X'$ component of the bulk velocity is largest
near $Z'=0$, while the toroidal-like Biermann magnetic field becomes weak and vanishes
on the $X'$ axis.
The Al plasma can therefore preferentially expand outward through this weak-field region
around $Z'=0$, producing a protruding rounded leading edge.
The characteristic size of this structure is related to the width of the weak-field
region, corresponding to the current-sheet region around the $X'$ axis.
At the injection boundary, this width is approximately $0.2R_{\rm in}\simeq0.04$~cm
[see Eq.~\ref{eq:B-Biermann}],
and it broadens as the Al plasma propagates outward, reaching approximately $0.2$~cm
at around 33~ns in the MHD simulation.
The MHD calculations also indicate that this scale depends on the density structure
of the ejecta and the plasma $\beta$.

In the experiment, the characteristic size of the rounded leading edge is
approximately $0.4$~cm, whereas in the MHD simulation it is about $0.2$~cm.
These scale lengths are estimated from the apparent size of the high-density apex
of the interface almost along the $X'$ axis,
using the experimental self-emission image [Fig.~\ref{fig:SE_33ns}(a) and (b)]
and the corresponding MHD density maps [Fig.~\ref{fig:MHD_33ns}(a) and (b)],
respectively.
The difference by a factor of about two is consistent with the systematically shorter
ripple wavelength obtained in the simulation.
Importantly, the rounded leading edge itself persists during the evolution rather than
being periodically generated and disappearing.
In the MHD simulation, ripple structures develop from the leading-edge region and
propagate backward along the interface.
The experimental data do not provide a continuous time sequence that would allow this
propagation to be followed directly; nevertheless, the gradual increase of the ripple
wavelength with time is qualitatively similar to that found in the simulation.
This suggests that the ripple structures observed in the experiment and simulation
have the same physical origin.
Furthermore, the formation of the rounded leading edge does not require an initially
curved piston.
Even if the injected Al plasma were initially planar, preferential expansion through
the weak-field region around the $X'$ axis would produce a local protrusion.
Thus, the formation of the leading edge and the resulting ripple structure are not
essentially determined by the initial curvature radius of the piston.

We note that velocity shear is present around the rounded leading edge, where vorticity
is generated.
Therefore, we cannot exclude the possibility that a KH-type instability
develops locally in this region.
Such a local instability, however, would be expected to produce structures on scales
associated with the local shear layer and therefore shorter than the dominant ripple
wavelength considered here.
We therefore interpret the observed large-scale ripple structure primarily as being
associated with the vortical motions generated around the persistent rounded leading edge,
while local KH-type instability may contribute to the smaller-scale
development of the interface.
Reproducing the detailed three-dimensional structure of the leading edge in the present
two-dimensional, simplified MHD simulation is inherently difficult; nevertheless,
the simulation captures the essential features of the ripple formation and its spatial
scale in reasonable agreement with the experimental observations.

Although the shock in the ambient N plasma is clearly visible in the density maps in Fig.~\ref{fig:MHD_33ns}(a) and (b) (e.g., along the $X'$ axis, the shock front appears around $X' \approx 2.2$--$2.3~\mathrm{cm}$), it is not prominent in the synthetic self-emission maps in Fig.~\ref{fig:MHD_33ns}(e) and (f). 
A similar situation is also seen in the experimental GOI images [Figs.~\ref{fig:SE_By-} and \ref{fig:SE_33ns}(a) and (b)]: 
for both the experimental results and the MHD simulations, the brightest self-emission arises at the interface, where the emission is dominated by the Al plasma. The main reason is the comparatively low electron density in the N plasma, which makes its bremsstrahlung emissivity much smaller than that of the dense interface region. In addition, the self-emission from shock-compressed N plasma can even become weaker downstream than upstream, because the bremsstrahlung emissivity, given by Eq.~(\ref{eq:bremss_MHD}), 
decreases as a result of the substantial temperature increase behind the shock, and the resulting reduction can compensate for the density increase.

Finally, we stress that the shock transition layer in the ambient N plasma is expected to have a thickness of order the ion inertial length; for the upstream N plasma with $Z_{\rm N}=3$, this corresponds to $c/\omega_{pi} \sim 0.1~\mathrm{cm}$. This implies that an MHD description is not applicable for the shock in the N plasma, and we therefore refrain from further MHD-based discussion of the shock structure here. An analysis resolving the shock transition layer and ion-scale effects is presented in the companion paper \cite{Yamazaki2025a}, while a brief discussion using the self-emission profiles (Fig.~\ref{fig:SE_Xslice}) is given in Sec.~\ref{subsec:iondynamics}.

\section{Discussion}
\label{sec:discuss}

\subsection{Origin of density asymmetry in the Al plasma}
\label{subsec:densityasym}

So far, there have been no reports indicating that
an external magnetic field affects the ejection of laser-ablation plasma.
Our MHD simulations show that the observed polarity-dependent asymmetry can be qualitatively reproduced
when the Al plasma injected from the target side is assumed to have a density asymmetry whose direction depends on the polarity of the external magnetic field [Eqs.~(\ref{eq:rho_B0_posi}) and (\ref{eq:rho_B0_nega})].
Here, we briefly discuss a possible physical origin of such an initial asymmetry.

When the drive laser first reaches the Al target, the laser-induced force, mainly the ponderomotive force, acts on the electrons approximately along the laser-propagation direction, i.e., the $-X'$ direction.
The resulting displacement of electrons relative to ions produces an initial charge separation along the $X'$ direction.
At this early stage, the self-generated Biermann magnetic field in the emerging Al plasma may not yet be fully developed.
As the laser continues to deposit energy into the target, the electrons are therefore subject to both the laser-induced force in the $-X'$ direction and the externally applied magnetic field in the $y$ direction.
The resulting force drift of the electrons (i.e., $\vec{F}\times\vec{B}$ drift) is toward $+Z'$ for $B_{y0}>0$ and toward $-Z'$ for $B_{y0}<0$.
Because the ions cannot instantaneously follow this transverse electron drift, an additional charge separation develops in the $Z'$ direction, producing an electrostatic-field component that accelerates the Al ions in the same transverse direction as the electrons.
This process may provide the origin of the polarity-dependent asymmetry generated during laser irradiation, with enhanced Al-plasma ejection toward $+Z'$ for $B_{y0}>0$ and toward $-Z'$ for $B_{y0}<0$.
These directions are consistent with the asymmetric Al-plasma injection assumed in Eqs.~(\ref{eq:rho_B0_posi}) and (\ref{eq:rho_B0_nega}).

In reality, the laser-ablation process involves additional effects, including the subsequent growth of the self-generated Biermann magnetic field, pressure gradients, radiation from the Al plasma, electrostatic fields, collisions, and multidimensional plasma expansion.
A quantitative determination of the origin of the observed asymmetry therefore requires a more self-consistent treatment of the laser--target interaction and is left for future work.

\subsection{Ion dynamics around the interface}
\label{subsec:iondynamics}

In the present experiment, we identified 
the rippling electron-scale interface between the Al and N plasmas.
Outside the boundary (i.e., on the upstream side), the plasma is 
predominantly composed of tenuous N plasma, for which the fluid approximation 
does not necessarily hold for ions and their kinematic behavior is important.
Such an electron-scale interface and the associated ion dynamics were already demonstrated in the higher-density ($P_{\rm N}=5$~Torr) 
experiments \cite{Yamazaki2022} and PIC simulations \cite{Umeda2019}, 
which showed Al-ion penetration into the N plasma.
Under the present experimental condition ($P_{\rm N}=1$~Torr), 
the N plasma is even more 
rarefied, and the collisionless processes between the Al and N plasmas become
more pronounced.
Indeed, using a representative relative velocity of 500~km~s$^{-1}$, 
the interspecies ion--ion Coulomb mean free paths, calculated in the same manner as in Ref.~\cite{Yamazaki2022}, 
are $\gtrsim0.5$~cm in both directions, for N ions penetrating into the Al plasma and for Al ions penetrating into the N plasma
\footnote{
For Al ions penetrating into the N plasma at a relative velocity of 500~km~s$^{-1}$, 
we adopt $Z_{\rm N}=6$, $Z_{\rm Al}=11$, $T_{\rm e}=T_{\rm i}=100$~eV, and 
$n_{\rm i}=1.28\times10^{17}$~cm$^{-3}$, where the ion density is assumed to be four times that of the far-upstream N plasma. 
These parameters give $\lambda_{\rm ii}=1.3$~cm. 
Assuming $B_{\rm N}=10$~T, the Al-ion gyro-radius is $r_{\rm g,Al}=0.13$~cm, yielding 
$\lambda_{\rm ii}/r_{\rm g,Al}=10.2$. 
Even for $T_{\rm e}=T_{\rm i}=300$~eV, the ratio remains $\lambda_{\rm ii}/r_{\rm g,Al}=9.8$. 
For N ions penetrating into the Al plasma at the same relative velocity, we use the plasma parameters inferred from the Thomson-scattering measurements described in Appendix~A: 
$Z_{\rm N}=6$, $Z_{\rm Al}=11$, $T_{\rm e}=377$~eV, $T_{\rm i}=882$~eV, and 
$n_{\rm i}=1.8\times10^{17}$~cm$^{-3}$. 
We then obtain $\lambda_{\rm ii}=0.49$~cm. 
Assuming $B_{\rm Al}=10$~T gives $r_{\rm g,N}=0.13$~cm and 
$\lambda_{\rm ii}/r_{\rm g,N}=4.0$
}.
These values are larger than the observed transition scale ($\lesssim0.1$~cm) 
and are also longer than the corresponding ion gyro-radii, indicating that ion--ion Coulomb scattering is unlikely to determine the narrow interface width.
The ion gyro-radii themselves are of the order of 0.1~cm.
On the other hand, as discussed in Ref.~\cite{Yamazaki2022}, 
the apparent width of a microscopic interface in the self-emission image can be broadened by line-of-sight integration 
and the three-dimensional curvature of the interface, i.e., by projection effects.
Therefore, the similar apparent transition widths observed with and without the external magnetic field do not necessarily imply that the interface width is controlled by ion--ion Coulomb collisions.
Taking these projection effects into account, the observed narrow transition is consistent with the interpretation of an electron-scale Al--N interface, although its intrinsic microscopic thickness is not directly resolved by the present imaging diagnostic.
Based on this physical picture, we interpret the 
self-emission profiles shown in Fig.~\ref{fig:SE_Xslice} in terms of the ion dynamics around the Al--N interface.

In Fig.~\ref{fig:SE_Xslice}, all four curves corresponding to 
$B_{y0}=\pm4$~T, $B_{y0}=0$ with $P_{\rm N}=1$~Torr, and $B_{y0}=P_{\rm N}=0$ 
almost coincide in the region $X'<1.0$~cm, indicating that this inner region 
is mainly occupied by the Al plasma.
For $X'>1.3$~cm, one can see differences among the three cases with $P_{\rm N}=1$~Torr.

Without the external magnetic field ($B_{y0}=0$) but with N gas ($P_{\rm N}=1$~Torr), 
Al ions pass through the electron-scale interface and penetrate into the upstream N plasma because the N plasma 
carries no magnetic field \cite{Umeda2019}.
Since Al ions cannot push the N plasma strongly, the electron density does not become sufficiently large, 
and the intensity peak of the self-emission appears at a position slightly inside the interface.
The upstream N plasma is heated by the penetrating Al ions 
\cite{Umeda2019,Yamazaki2022}, 
producing a somewhat brighter emission than in the magnetized cases.
This interpretation is also supported by the GOI images. 
In the unmagnetized case, a faint and diffuse emission component is observed farther upstream 
[e.g., around $X'\approx2.75$~cm and $Z'\approx0.75$~cm in Fig.~\ref{fig:SE_33ns}(c)]. 
Similar features are also observed at earlier times, and their propagation suggests a characteristic speed of approximately 800--1000~km~s$^{-1}$. 
We interpret this emission as being associated with the leading, fast Al ions penetrating into the ambient N plasma, consistent with our previous experiment \cite{Yamazaki2022}. 
Because the emission is weak and diffuse, its propagation speed cannot be determined accurately. 
Such deep penetration of the Al ions broadens the Al--N interaction region. 
Consequently, within the temporal and spatial ranges accessible in the present experiment, a shock forming in the N plasma cannot be clearly separated from the broad Al--N interaction region in the unmagnetized case.

With the external magnetic field ($B_{y0}=\pm4$~T), by contrast, 
Al ions attempting to cross the interface are turned back by the strong 
magnetic field embedded in the swept-up N plasma and cannot propagate 
further upstream \cite{Umeda2019,Yamazaki2022}. 
Because Al ions are 
prevented from entering the N plasma, they accumulate near the interface, 
leading to enhanced plasma density and producing a sharp and pronounced 
emission peak at the interface. In the upstream N-plasma region, the emission 
intensity gradually decreases with distance, reflecting the presence of 
reflected N ions in that region.
That structure is the foot-like region of a forming supercritical magnetized shock.
Further details are discussed in the companion paper \cite{Yamazaki2025a}.

\section{Conclusion}
\label{sec:conclusion}

We have performed laser--plasma interaction experiments on the early stage of the formation of quasi-perpendicular supercritical magnetized collisionless shocks
in a low-pressure ambient N gas
($P_{\rm N}=1~{\rm Torr}$) under an externally applied magnetic field with controlled polarity,
and investigated the evolution of self-emission structures formed by the interaction between
the expanding Al plasma and the magnetized N plasma.
Self-emission imaging revealed a bright outer emission front that propagates outward and develops
ripple-like modulations.
The global emission pattern reverses when the magnetic-field polarity is changed, whereas the
case without an external magnetic field exhibits weaker emission and a less distinct outer boundary.

Two-dimensional ideal-MHD simulations reproduce the main features of the observed emission
structures, including the bow-shaped interaction layer and the polarity-dependent asymmetry.
An interpretation based on the KH instability is disfavored, as the sonic Mach number
along the interface is of order unity and the estimated growth time exceeds the observed timescale
of ripple formation.
Instead, the ripples are more consistently explained by vortex formation associated with the
high-density, rounded leading edge of the emission front near the $X'$ axis, which sets the
characteristic ripple wavelength.

A key result of this study is that the bright, rippling emission delineates the electron-scale interface between
Al and N plasmas.
The region on its upstream side is predominantly composed of magnetized N plasma.
This allows the discussion of shock formation in the magnetized N plasma to be 
carried out clearly, with minimal influence from Al-plasma contamination.
The formation of the shock becomes evident around half of the gyro period of upstream N ions
with charge state $Z_{\rm N}\approx3$ \cite{Yamazaki2022},
$\pi/\omega_{ci}\approx 38$~ns, when the shock starts separating from the interface.
Together, the experimental diagnostics and MHD simulations provide a consistent picture of the
electron-scale interface and its role in shaping the ripple structure and the early evolution of
magnetized collisionless shocks.



\begin{acknowledgments}
The authors would like to acknowledge the dedicated technical
support of the staff at the Gekko-XII facility for the laser operation,
target fabrication, and plasma diagnostics.
This work was the result of using research equipment shared in MEXT Project for promoting public utilization of advanced research infrastructure
(Program for advanced research equipment platforms) Grant Number JPMXS04503001.
We would also like to thank 
N.~Ozaki for the target alignment in the experiment, 
and N.~Yamamoto, T.~Sakamoto, and H.~Maeda 
for useful discussions. 
Numerical computations were carried out on Cray XD2000 at the Center for Computational Astrophysics, National Astronomical Observatory of Japan.
This research was partially supported by 
the joint research project of Institute of Laser Engineering, Osaka University, 
JSPS KAKENHI Grants No. 
%
23K22522, 24K00605, 23K25907, 24H01802, 24H01816, 24K00603,
21H04487, 24H01805, 22K14020, 24K17029,
22H00119, 23H04864, and 25K00999,
Grant-in-Aid for JSPS Fellows No. 24KJ0034, 
the Kajima Foundation (International Collaborative Research) (S.M.),  
and
JSPS Core-to-Core Program B: Asia-Africa Science Plat-forms Grant No. JPJSCCB20190003 (Y.~Sakawa).
\end{acknowledgments}


\appendix

\section{Collective Thomson scattering (CTS) measurements}
\label{sec:TSmeasure_SOP}

In this Appendix, we briefly summarize the supplementary diagnostics employed in this study:
 streaked optical pyrometry (SOP) of the self-emission 
and  CTS measurement.
They provide complementary information on the plasma conditions associated with the self-emission
imaging recorded by GOI.
The SOP and CTS measurements indicate that the plasma density in the region of bright self-emission 
is $n_e \approx 2 \times 10^{18}$~cm$^{-3}$, corresponding to an ion density of
$n_i = n_e/Z_{\rm Al} \approx 1.8 \times 10^{17}$~cm$^{-3}$, where $Z_{\rm Al}$ is the charge state of the Al plasma.

The SOP technique measures the time evolution of the self-emission intensity
along the $X'$ axis.
The experimental configuration and measurement scheme of the SOP diagnostic
are identical to those  in the companion paper \cite{Yamazaki2025a},
and  follow those used in our previous works \cite{Yamazaki2022,Matsukiyo2022}.
The CTS diagnostics were used to infer plasma parameters through spectral analysis
of the scattered light. Details are described in the companion paper \cite{Yamazaki2025a}.
In the present study, the following two cases were analyzed.
\begin{itemize}
\item
$B_{y0} = -4$~T at $t = 33$~ns.
Under this condition, the CTS measurements targeted the ion feature.
The diagnostic geometry and analysis procedure are consistent with the ``IAW-1'' configuration
used in the companion paper \cite{Yamazaki2025a} and are identical to those employed in
our previous studies \cite{Yamazaki2022,Matsukiyo2022,Morita2019,Morita2022,Sakai2022}.
The SOP measurement in this case was performed using self-emission through 
a blue filter (central wavelength $450 \pm 2$~nm, FWHM $10 \pm 2$~nm).
\item
$B_{y0} = +2.7$~T at $t = 29$~ns.  
Under this condition, the CTS measurements focused on the electron feature.
This configuration is identical to the ``EPW'' setup described in the companion paper \cite{Yamazaki2025a}
and is also the same as that used in Matsukiyo \textit{et al.} \cite{Matsukiyo2026a}.
The corresponding SOP measurement was performed using self-emission through
a green filter (central wavelength $532 \pm 2$~nm, FWHM $10 \pm 2$~nm).
\end{itemize}
In the following subsections, we show the results for each case.
Note that the electron density of $n_e\approx2\times10^{18}$~cm$^{-3}$ derived in both cases is roughly consistent with
the value of the bright self-emission region $n_e\approx4\times10^{18}$~cm$^{-3}$  for the case of $P_{\rm N}=5$~Torr and  $B_{y0}=-3.6$~T 
at $X'=1.4$~cm and $t=30$~ns \cite{Yamazaki2022}.

\subsection{Results for $B_{y0}=-4$~T at $t=33$~ns}

First, we present the results for the case of $P_{\rm N} = 1$~Torr,
$B_{y0} = -4$~T at $t = 33$~ns.
Figure~\ref{fig:TSi33ns_TSe29ns}(a)--(c) summarize the corresponding SOP and CTS
measurements.

As shown by the MHD simulations, the brightest self-emission region near the
TCC at  $t \approx 33$~ns is dominated by Al plasma.
Accordingly, we assumed that the plasma responsible for the ion feature in the CTS spectrum was 
Al plasma and fitted the spectrum using the same procedure as in our previous work \cite{Yamazaki2022}.
Figure~\ref{fig:TSi33ns_TSe29ns}(c) shows the background-subtracted spectrum of
the CTS ion feature.
The measured spectrum exhibits a clear double-peak structure, which reflects
the resonance with ion-acoustic waves.
We thus obtained the following plasma parameters:
electron density $n_e = (1.97 \pm 0.26) \times 10^{18}$~cm$^{-3}$,
ion charge state $Z_{\rm Al} = 11$,
electron and ion temperatures $T_e = 377 \pm 19$~eV and
$T_i = 882 \pm 16$~eV, respectively,
and the $x$ component of the bulk flow velocity
$v_x = 456.1 \pm 0.5$~km~s$^{-1}$.

The observed  asymmetric double-peak structure in Fig.~\ref{fig:TSi33ns_TSe29ns}(c)
may be explained by 
 a relative drift between Al-ion and electron components \cite{sheffield2010plasma},
  by a deviation of the ion velocity distribution
function from a Maxwellian \cite{Morita2022}.
and/or by two-stream plasmas \cite{Sakai2023}.
In the present analysis, however, we assumed a Maxwellian ion distribution without the relative drift velocity;
consequently, the best-fit theoretical spectrum (the solid red curve) is symmetric,
while still reproducing the overall spectral features.

\subsection{Results for $B_{y0}=+2.7$~T at $t=29$~ns}

\begin{figure}[t]
\includegraphics[width=140mm, bb= 0 0 1200 900]{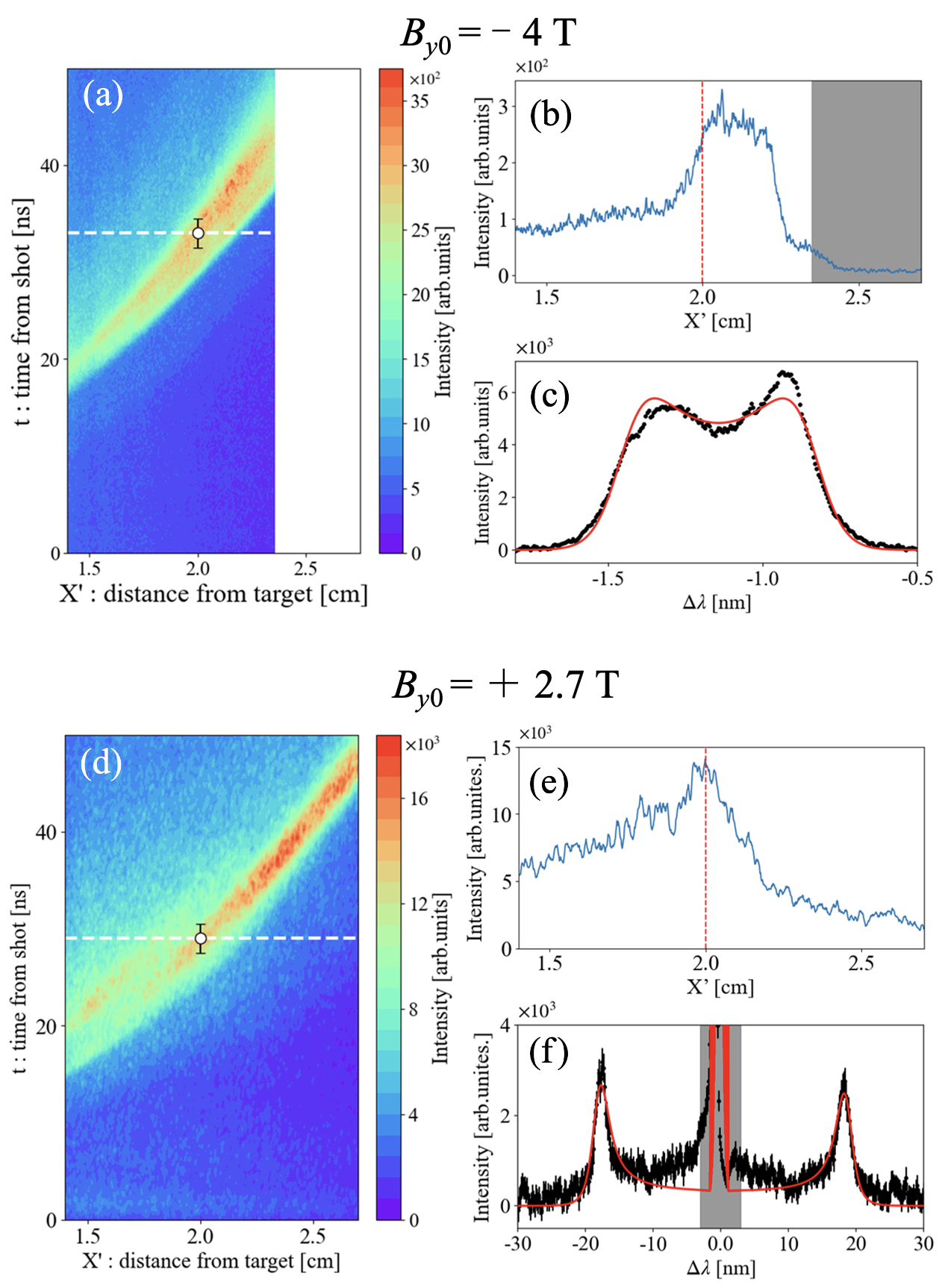}
\caption{
Results of streaked optical pyrometry (SOP) and collective Thomson scattering (CTS)
measurements for $P_{\rm N} = 1$~Torr. Panels (a)--(c) correspond to the case of
$B_{y0} = -4$~T, and panels (d)--(f) to the case of $B_{y0} = +2.7$~T.
Panels (a) and (d) show SOP images representing the time evolution of the
self-emission intensity along the $X'$ axis. 
The white circles indicate the times and positions of the CTS measurements:
(a) $t = 33 \pm 1.5$~ns and (d) $t = 29 \pm 1.5$~ns, 
both taken at the target chamber center (TCC; $X' = 2.0$~cm). 
Panels (b) and (e) show spatial profiles of the
self-emission intensity along the $X'$ axis at the times indicated by the white
dashed lines in panels (a) and (d), respectively. The vertical red dotted lines
mark the TCC position where the CTS spectra were measured.
The gray-shaded
region in panel~(b) indicates a region of  small detector gain. 
Panel (c) shows the background-subtracted
CTS spectrum (ion-feature) for $B_{y0} = -4$~T.
The black dots represent the experimental data, and the red curve is the
best-fit theoretical spectrum.
Panel (f) shows the background-subtracted CTS spectrum (electron feature) for $B_{y0} = +2.7$~T,
with the red curve representing the best-fit theoretical spectrum. 
The shaded region around the incident wavelength
($\lambda = 532$~nm, $\Delta\lambda = 0$) is affected by stray light.
\label{fig:TSi33ns_TSe29ns}
}
\end{figure}

We next present the results for the case of $P_{\rm N} = 1$~Torr and $B_{y0} = +2.7$~T at $t = 29$~ns.
Panels~(d)--(f) in Fig.~\ref{fig:TSi33ns_TSe29ns} show the corresponding SOP and CTS
measurements.
The SOP image [panel~(d)] and the self-emission intensity
profile along the $X'$ axis at $t = 29$~ns [panel~(e)]
indicate that the Thomson-scattering measurement probes the bright plasma region
around the TCC.
In this case, we obtained the electron feature of the CTS spectrum, whose shape is
determined by the electron
density $n_e$ and electron temperature $T_e$. 
We fitted the spectrum shown in
Fig.~\ref{fig:TSi33ns_TSe29ns}(f) to infer these parameters and obtained
$n_e = (2.61 \pm 0.02) \times 10^{18}$~cm$^{-3}$ and
$T_e = 261 \pm 4$~eV.
Assuming a collisional--radiative model for Al plasma under these conditions \cite{Chung2005},
the corresponding mean charge state is estimated to be
$Z_{\rm Al} \approx 11$.

The measured electron density of $n_e\approx2.6\times10^{18}$~cm$^{-3}$ is so high 
 that the bright peak of the plasma self-emission does not contain 
a significant contribution from shock-compressed downstream N plasma.
Let us assume that both the self-emission and CTS scattered light around $X'=2.0$~cm at $t=29$~ns  
originated from the shock-compressed N plasma.
 Because the far-upstream N plasma has $n_e=0.64Z_{\rm N}\times10^{17}$~cm$^{-3}$, where $Z_{\rm N}$ is the charge state of the
N plasma ($Z_{\rm N}\leq7$), the downstream electron density would be at most $n_e=1.8\times10^{18}$~cm$^{-3}$, 
assuming a compression ratio of four.
This value is smaller than the measured one, $n_e\approx2.6\times10^{18}$~cm$^{-3}$. 
Hence, the measured electron density cannot be explained by the shock-compressed N plasma.
This also strengthens our conclusion that the region of bright self-emission is dominated by Al plasma.

\bibliographystyle{apsrev4-2}
\bibliography{paper2}


\end{document}